\documentclass[sigconf]{acmart}

\usepackage{microtype}
\usepackage{graphicx}
\usepackage{subfigure}
\usepackage{float}
\usepackage{booktabs} 

\usepackage{amsmath}               
\usepackage[ruled,vlined]{algorithm2e}
\usepackage{setspace}
\usepackage{multirow}
\usepackage{enumitem}

\usepackage{xcolor}
\usepackage{titlesec}

\theoremstyle{definition}
\newtheorem{definition}{Definition}[section]

\theoremstyle{Assumption}
\newtheorem{assumption}{Assumption}[section]

\theoremstyle{Lemma}
\newtheorem{lemma}{Lemma}[section]

\theoremstyle{Theorem}
\newtheorem{theorem}{Theorem}[section]

\usepackage{hyperref}

\usepackage{cleveref}

\usepackage{comment}

\def\model{FedAT} 

\AtBeginDocument{%
  \providecommand\BibTeX{{%
    \normalfont B\kern-0.5em{\scshape i\kern-0.25em b}\kern-0.8em\TeX}}}

\renewcommand\footnotetextcopyrightpermission[1]{} 
\begin{document}

\title{FedAT: A High-Performance and Communication-Efficient Federated Learning System with Asynchronous Tiers}



\author{Zheng Chai}
\affiliation{%
  \institution{George Mason University}
  \streetaddress{Anonymous}
  \city{Fairfax, VA}
  \country{USA}}
\email{zchai2@gmu.edu}

\author{Yujing Chen}
\affiliation{%
  \institution{George Mason University}
  \streetaddress{Anonymous}
  \city{Fairfax, VA}
  \country{USA}}
\email{ychen37@gmu.edu}

\author{Ali Anwar}
\affiliation{%
  \institution{IBM Research - Almaden}
  \streetaddress{Anonymous}
  \city{San Jose, CA}
  \country{USA}}
\email{ali.anwar2@ibm.com}

\author{Liang Zhao}
\affiliation{%
  \institution{Emory University}
  \streetaddress{Anonymous}
  \city{Atlanta, GA}
  \country{USA}}
\email{liang.zhao@emory.edu}

\author{Yue Cheng}
\affiliation{%
  \institution{George Mason University}
  \streetaddress{Anonymous}
  \city{Fairfax, VA}
  \country{USA}}
\email{yuecheng@gmu.edu}

\author{Huzefa Rangwala}
\affiliation{%
  \institution{George Mason University}
  \streetaddress{Anonymous}
  \city{Fairfax, VA}
  \country{USA}}
\email{rangwala@gmu.edu}

\settopmatter{printacmref=false}
\settopmatter{printfolios=true}


\begin{abstract}
Federated learning (FL) involves training a model over massive distributed devices, while keeping the training data localized and private. This form of collaborative learning exposes new tradeoffs among model convergence speed, model accuracy, balance across clients, and communication cost, with new challenges including: (1) straggler problem---where clients lag due to data or (computing and network) resource heterogeneity, and (2) communication bottleneck---where a large number of clients communicate their local updates to a central server and bottleneck the server. Many existing FL methods focus on optimizing along only one single dimension of the tradeoff space. Existing solutions use asynchronous model updating or tiering-based, synchronous mechanisms to tackle the straggler problem. However, asynchronous methods can easily create a communication bottleneck, while tiering may introduce biases that favor faster tiers with shorter response latencies. 

To address these issues, we present \textbf{\model}, a novel \textbf{Fed}erated learning system with \textbf{A}synchronous \textbf{T}iers under Non-i.i.d. training data. {\model} synergistically combines synchronous, intra-tier training and asynchronous, cross-tier training. 
By bridging the synchronous and asynchronous training through tiering, {\model} minimizes the straggler effect with improved convergence speed and test accuracy. {\model} uses a straggler-aware, weighted aggregation heuristic to steer and balance the training across clients for further accuracy improvement.  {\model} compresses uplink and downlink communications using an efficient, polyline-encoding-based compression algorithm, which minimizes the communication cost. Results show that {\model} improves the prediction performance by up to $21.09\%$ and reduces the communication cost by up to $8.5\times$, compared to state-of-the-art FL methods. 
\end{abstract}




\maketitle
\pagestyle{plain} 

\section{Introduction}
The number of intelligent devices, such as smartphones 
and wearable devices, has been rapidly growing in the last few years. Many of these devices are equipped with various smart sensors and increasingly potent hardware that allow them to collect and process data at unprecedented scales. With advanced machine learning techniques and the growth in computation power of these devices, federated learning (FL) has emerged as a novel machine learning paradigm that aims to train a statistical model among a large number of edge device nodes (clients\footnote{We use ``clients'' and ``devices'' interchangeably in the paper.}), as opposed to traditional machine learning training at a centralized location~\cite{mcmahan2017communication,konevcny2016federated}. 
FL has been used in many application domains, including predicting human activities~\cite{chen2019communication,chen2019asynchronous}, learning sentiment~\cite{smith2017federated}, language processing~\cite{li2019fair,yang2018applied,hard2018federated}, and enterprise infrastructures~\cite{ludwig2020ibm}.

In a typical FL framework, a shared model is learned from a federation of distributed clients with the coordination of a server, and clients do not share data with each other due to security and privacy reasons~\cite{tankard2016gdpr, o2004health}. Each client trains a local model using its (decentralized) local data. 

FL often involves a large number of clients, which feature highly heterogeneous hardware resources (CPU, memory, and network resources) and \textbf{Non-i.i.d.} (non-independent and identical) data; that is, the training data distributed across the clients is often non-uniform, since the data generated by a given client is typically based on the usage of that particular edge device and would not be representative of the overall population distribution~\cite{mcmahan2017communication,li2019convergence,sattler2019robust,zhao2018federated}. The resource and data heterogeneity present unique challenges to FL algorithms. In addition, with large number of clients, how clients communicate with server becomes a crucial design choice. Most existing FL frameworks can be divided into two communication modes: (1) synchronous communication (e.g., Federated Averaging, or FedAvg~\cite{mcmahan2017communication}), or (2) asynchronous communication (e.g., FedAsync~\cite{xie2019asynchronous}). When there are stragglers (i.e., slower clients) in the system, which is common especially at the scale of hundreds of clients, asynchronous approaches are more robust. However, most asynchronous implementations suffer communication bottleneck as all the clients can asynchronously talk to the server, and clients are limited by their communication bandwidth. Therefore, in this paper, we focus on two significant challenges of federated learning: 

\noindent\textbf{Stragglers.} 
Recent research efforts in synchronous FL assume: i) no resource or data heterogeneity~\cite{reisizadeh2020fedpaq,sattler2019robust}, or ii) all the clients are available during the whole training process~\cite{zhao2018federated,nishio2019client}. However, in practice, clients may come (online) and go (offline) frequently, or lag due to resource/data heterogeneity (i.e., stragglers). Existing synchronous FL solutions (e.g., Federated Averaging, or FedAvg~\cite{mcmahan2017communication}) synchronously aggregates model updates, where the server has to wait for the slowest clients,  therefore, leading to significantly prolonged training time. 

\noindent\textbf{Communication Bottleneck.}
To solve the straggler problems in synchronous FL, asynchronous FL approaches~\cite{chen2019asynchronous,xie2019asynchronous} were proposed, where the server can aggregate without waiting for the straggling clients. Unlike synchronous FL where only a portion of sampled clients communicate with the server in each training round, in an asynchronous FL setting, the server communicates with all the clients asynchronously; therefore, the server can easily become a communication bottleneck with tens of thousands of clients updating the model simultaneously.

To overcome the deficiencies described above, we design and implement {\model}, a novel communication-efficient FL approach that combines the best of both worlds -- synchronous and asynchronous FL training -- using a tiering mechanism. 
In {\model}, the clients involved in a FL deployment are partitioned into logical tiers based on their response latency (i.e., the time a client takes to finish a single round of training). 
All the logical tiers in {\model} participate in the global training simultaneously, with each tier proceeding at its own pace. Clients within a single tier update a model associated with that particular tier in a synchronous manner, 
while each tier, as a logical, coarse-grained, training entity, updates a global model asynchronously. Faster tiers, with shorter per-round response latency, drive the global model training to converge faster; slower tiers get involved in the global training by asynchronously sending the model updates to the server, so as to further improve the model's prediction performance.

Uniformly aggregating the asynchronously updated tier model into the global model may result in biased training (biased towards the faster tiers), as more performant tiers tend to update the global model more frequently than the slower tiers. 
To solve this issue, we propose a new weighted aggregation heuristic,
which assign higher weight to slower tiers. Furthermore, to minimize the communication cost introduced by asynchronous training, {\model} compresses the model data transferred between the clients and the server using Encoded Polyline Algorithm. 
In a nutshell, {\model} synergizes the four components together, namely, the tiering scheme, asynchronous inter-tier model updating, the weighted aggregation method, and polyline encoding compression algorithm, to maximize both the convergence speed and the prediction performance while minimizing communication cost.

We make the following contributions in this paper:
\begin{itemize}[noitemsep,leftmargin=*]
    \item We design and implement {\model}, a novel, tiered, FL framework, which updates local model parameters synchronously within tiers and updates the global model asynchronously across tiers.
    \item We propose a new optimization objective with a weighted aggregation heuristic, which the FL server uses to speed up the model convergence and improve the prediction performance by balancing the model parameters from different tiers. 
    \item We provide rigorous, theoretical analysis for our proposed method for both convex and non-convex objectives; our analysis shows that {\model} has provable convergence guarantee.
    \item We utilize a lossy compression technique---polyline encoding---to compress the transferred model data between clients and server to reduce the communication cost without affecting the model accuracy.
    \item We evaluate {\model} extensively on a medium-scale, 100-client cluster on Chameleon Cloud and a large-scale, 500-client cluster on AWS EC2. Experimental results on five federated datasets including CIFAR-10, Fashion-MNIST, Sentiment140, FEMNIST, and Reddit under an FL benchmarking framework \textbf{LEAF}~\cite{caldas2018leaf} show that {\model} improves the prediction accuracy by up to $21.09\%$, exhibits significantly less accuracy variance during the training, and reduces the communication cost by up to $8.5\times$ compared to FedAsync~\cite{xie2019asynchronous}.
\end{itemize}

\section{Related Work}
\label{sec:related}

\subsection{Stragglers in Federated Learning}\label{sec:stragglers}

The main premise of FL has been collective learning using a network of computing devices such as smartphones and tablets. In such a training environment that cannot be fully controlled, data heterogeneity and resource heterogeneity may cause stragglers~\cite{chai2019towards}, which commonly exist in large-scale FL training scenarios~\cite{fmtl_nips17, flanp_arxiv, chai2020tifl}. Furthermore, in real-world scenarios, these clients could be frequently offline due to (computing/network) resource constrains. The assumption made by FedAvg that all clients are available during the whole training process is not practical. 

\noindent\textbf{Synchronous FL Frameworks}. 
Li et al.~\cite{li2019convergence} suggest to select a smaller ratio of clients for training in each global iteration to alleviate the straggler’s effect, while with more rounds for model convergence. However our experiments in \cref{sec:part_level} show that selecting less number of clients for each round produces lower performance. 
Bonawitz et al.~\cite{bonawitz2019towards} proposes a naive clients selection strategy to mitigate stragglers, where $130\%$ target number of clients are selected for each round. With this approach, the slowest $30\%$ is neglected. However, it comes with more communication cost and potential failure in handling stragglers when the number of stragglers involved in some rounds exceed the $30\%$ tolerance.
FedProx~\cite{li2018federated} tackles system heterogeneity by using distinct local epoch numbers for clients. However, choosing a perfect local epoch number for each client is challenging in real-world applications. 

TiFL~\cite{chai2020tifl} is a tier based FL framework that uses a synchronous, intra-tier model updating scheme similar to that used in FedAvg. The adaptive tier selection algorithm that TiFL relies on requires collecting test accuracies of all clients every certain rounds. This means higher communication costs and longer training duration. Our experiments results in \cref{sec:eval} show that this algorithm fails to 
achieve efficient communication given a high level of Non-i.i.d.-ness with larger number of  clients and may result in biased training and lower accuracy. 
In real-world scenarios, however, it is possible that a portion of clients are incorrectly profiled and assigned to a wrong tier as clients' response latencies may largely vary from time to time. {\model} uses TiFL's tiering approach but differs from TiFL in that {\model} combines intra-tier synchronous training with cross-tier asynchronous training to effectively mitigate training biases. With our new asynchronous mechanism, {\model} can tolerate mis-tiering caused by mis-profiling and performance variation.

\noindent\textbf{Asynchronous FL Frameworks}. 
Asynchronism is widely used in distributed systems for shortening overall running time~\cite{mnih2016asynchronous, fan2020adaptive, lian2018asynchronous,han2015giraph}. Asynchronous FL frameworks~\cite{xie2019asynchronous,chen2019asynchronous,lu2019differentially} allow wait-free communication and computation in order to address the straggler problem. These asynchronous approaches, however, suffer from high communication costs as they require much more frequent communications between clients and the server. Worse, frequent aggregation on the server side may lead to slow convergence speed. 
\vspace{-2em}
\subsection{Communication-Efficient Federated Optimization} 
 
McMahan et al. propose a FL approach called FedAvg~\cite{mcmahan2017communication}, where instead of communicating after every iteration, each client performs multiple iterations of SGD to compute a weight update. By reducing the communication frequency, FedAvg reduces the communication cost and can work with partial client participation. In a follow-up work, Kone{\v{c}}n{\`y} et al.~\cite{konevcny2016federated} propose two approaches to reduce the uplink communication costs, i.e., structured updates and sketched updates, combined with probabilistic quantization. 
These two approaches, however, are only suitable for i.i.d. settings in FL. For Non-i.i.d. settings, they can significantly slow down the convergence speed in terms of SGD iterations.

In the broader realm of communication-efficient distributed deep learning, a wide variety of methods has been proposed to reduce the amount of communication during the training process.
Chen et al.~\cite{chen2019communication} propose a layerwise asynchronous update scheme that updates the parameters of the deep layers less frequently than those of the shallow layers. Mills et al.~\cite{mills2019communication} adapt FedAvg to use a distributed form of Adam optimization and compress the uploaded parameters. 
Jeong et al.~\cite{jeong2018communication} develop federated distillation that follows an online version of knowledge distillation to compress the model.
Reisizadeh et al.~\cite{reisizadeh2020fedpaq} present a periodic averaging and quantization approach to reduce communication costs. However, these works only target uplink communication compression and are developed for synchronous frameworks that neglect the real-world scenario where stragglers are common. 
In addition to the above mentioned server-client topology, solving communication bottlenecks via quantization and compression has also gained considerable attention in decentralized training~\cite{koloskova2019decentralized,wang2019matcha,reisizadeh2019exact}. While such techniques can be used to reduce communication costs in FL, a decentralized network topology in distributed learning without a server is fundamentally different and is thus orthogonal to our approach.

Our proposed communication-efficient federated learning framework combines synchronous and asynchronous updates together to mitigate the challenge associated with stragglers and improve model convergence rate. We apply a weighted aggregation strategy on server to improve the model's prediction performance and compress both the uplink and downlink communications.

\section{Preliminaries: Federated Learning and FedAvg}\label{preliminaries}

FL algorithms involve hundreds to millions of remote devices training locally on their device-generated data, and collectively train a global, shared model, under the coordination of a centralized server serving as an aggregator.
In particular, the FL algorithm optimizes the following objective function:
\begin{equation}\label{syn-obj}
f(w) = \sum_{k=1}^{K}\frac{n_k}{N}F_k(w) ,
\end{equation}
where $F_k(w)\overset{def}= \frac{1}{n_k}\sum_{i\in\mathcal{D}_k}^{} \ell_i(x_i,y_i;w)$, is the local empirical loss of client $k$, and $\ell_i(x_i,y_i;w)$ is the corresponding loss function for data sample $\{x_i,y_i\}$. $K$ is the total number of devices.
$\mathcal{D}_k$ for $k \in \{1,\dots,K\}$ denotes data samples stored locally on device $k$. 
$n_k = |\mathcal{D}_k|$, is the number of data samples on device $k$; 
and $N = \sum_{k=1}^K |\mathcal{D}_k|$ is the total number of data samples stored on $K$ devices. 
Assuming for any $k\neq k^{'}$, $\mathcal{D}_k\bigcap\mathcal{D}_{k^{'}} = \varnothing$. 

The ultimate goal is to find a model $w_*$ that minimizes the objective function:
\begin{equation}\label{goal}
w_* = \arg \min f(w).
\end{equation}

\begin{algorithm}
\caption{Federated Averaging Training Algorithm}\label{alg:fedavg}
\textbf{Server:} Initialize global weights $w^0$\\
\For{each round $t=0$ to $T-1$}{
$\mathcal{S} = $ (random set of clients)\\
\For{each client $k$ $\in$ $\mathcal{S}$ \textbf{in parallel}}{
$w_k^{t+1} = w_k^t - \eta \nabla h(w^t)$
}
$N = \sum_{k=1}^{|\mathcal{S}|} n_k$\\
$w^{t+1} = \sum_{k=1}^{|\mathcal{S}|} \frac{n_k}{N} \cdot w_k^{t+1}$\\
}
\end{algorithm}
FedAvg~\cite{mcmahan2017communication} is a commonly used method to solve the optimization problem defined in Equation~\ref{goal} in a non-convex setting with a synchronous update fashion. This method runs by randomly sampling a subset of clients with a certain probability at each round; each selected client performs $E$ epochs of training locally on its own data using an optimizer such as stochastic gradient descent (SGD). The detailed process of FedAvg is shown in Algorithm~\ref{alg:fedavg}.
This kind of local update methods enable flexible and efficient communication compared to traditional mini-batch methods~\cite{yu2018parallel,wang2018cooperative,woodworth2018graph}.

In a typical real-world scenario, the data stored across devices follow a non-i.i.d. distribution.
Although FedAvg can work with partial client participation at each training
round, training on Non-i.i.d. data may lead each client towards its local optimal model as opposed to achieving a global optimal one.

In addition, slow clients (stragglers), which perform local training at a relatively slower speed (due to weaker computing resources and/or larger local data size), may have poor prediction performance due to less training, and thus may prevent the shared model from converging to a global optimal solution. 
Therefore, solving Equation~\ref{goal} in this synchronous manner can implicitly introduce high variance in prediction performance given the existence of straggling clients. To better evaluate the robustness of FL training with stragglers, we define new metrics as a measure of the straggler tolerance level in a typical FL setup.

\begin{definition}{(\textit{Robust training with straggling clients}).}\label{defi-straggle}
For two trained models $w$ and $w'$, we say that model $w$ is more robust against straggling clients than model $w'$, if (1) model $w$ \textit{converges faster} than model $w'$; 
(2) the test accuracy of model $w$ for the $K$ clients, $\{P_1,...,P_K\}$, exhibits \textit{lower variance} than that of model $w'$ for the same set of $K$ clients 
(where $P_c$ represents the test accuracy of the model $w$ over the testing data for client $c$), 
and (3) model $w$ has better prediction performance than model $w'$.  
\end{definition}

As discussed in the previous sections, the existence of stragglers causes longer training times and prevent the model from converging to the optimal solution. 
The rational of using these three metrics, namely, convergence speed, accuracy variance, and prediction accuracy, as a means to quantify the robustness of a FL approach against stragglers, is that:
(1) stragglers not only cause slow convergence speed, but also a loss in prediction performance,
and (2) existing literature fail to comprehensively consider all these aspects~\cite{chai2020tifl, chen2019communication, jeong2018communication, xie2019asynchronous,chen2019asynchronous}.

\begin{figure}[t]
\centering
\includegraphics[trim={8.5cm 0cm 1.8cm 0cm}, clip,width=8.5cm]{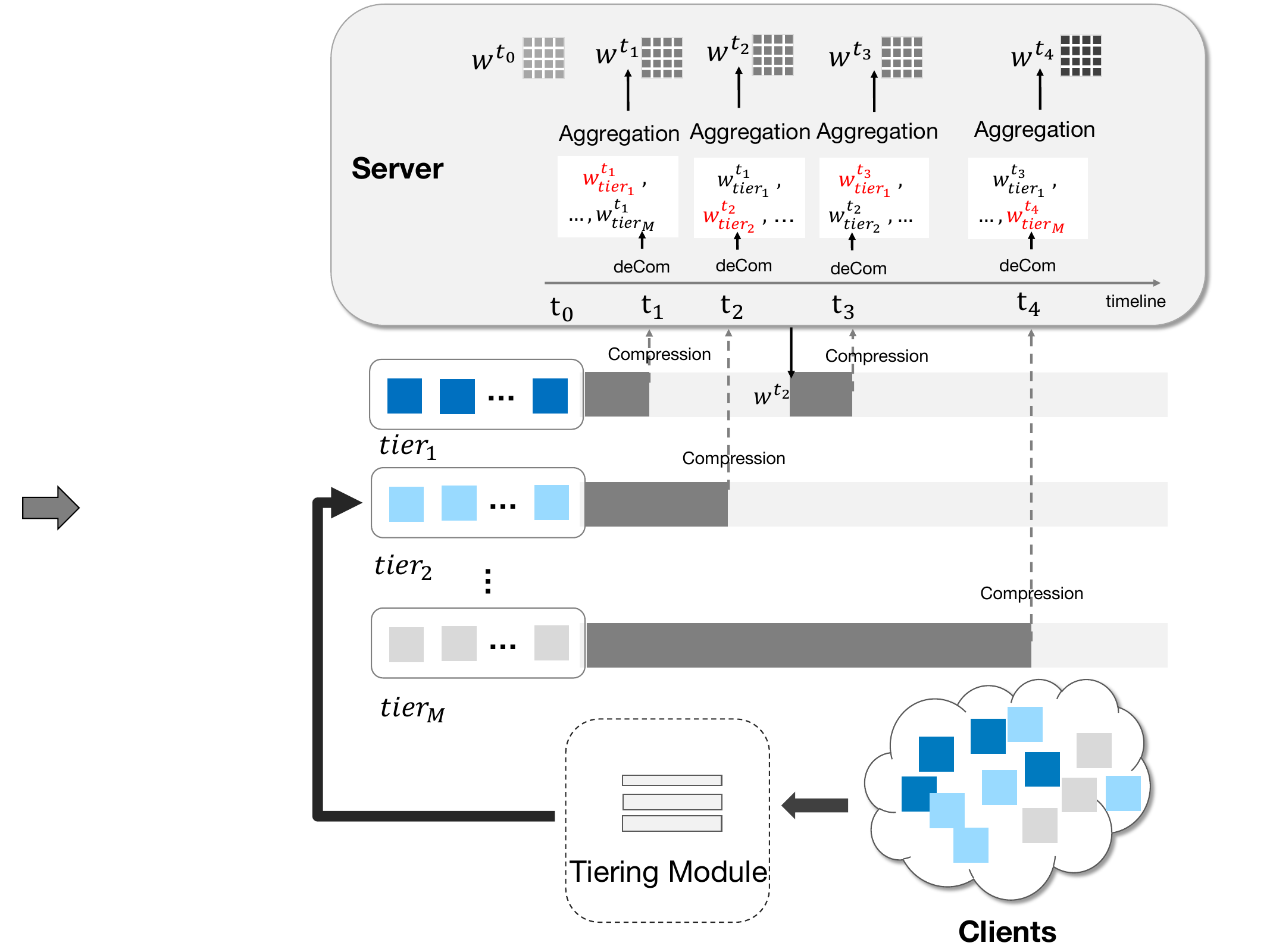}
	\caption{Overview of FedAT. $tier_1, ... , tier_M$ are $M$ tiers, and 
	$w_{tier_1}^t,...,w_{tier_M}^t$ are their according weights, respectively. 
	{\small\texttt{deCom}} denotes the decompression process of clients' models in a certain tier on the server. 
	}\label{system}
	\vspace{-5pt}
\end{figure}

\section{{\model}: Federated Learning with Asynchronous Tiers}

{\model} consists of three main components: 
(1) a centralized server for global model synchronization;
(2) a group of clients that are logically partitioned into different performance tiers;
and (3) a tiering module that profiles clients' training performance and performs client tiering based on the response latency of each client.

We next illustrate the FL training process of {\model} (as depicted in Figure \ref{system} and listed in Algorithm~\ref{alg:FedAT}). The tiering module profiles and partitions the involved clients into $M$ tiers based on their response latencies: \{$tier_1,tier_2,...,tier_M$\}, where $tier_1$ is the fastest tier and $tier_M$ is the slowest tier. The server maintains a list of $M$ models, \{$w_{tier_1}^t,...,w_{tier_M}^t$\}, one for each tier, reflecting the most updated view of per-tier local models, at a certain round $t$. Correspondingly, the server also maintains a global model $w$ that gets asynchronously updated from $M$ tiers. 

Each tier performs synchronous update process, a fraction of clients ($\mathcal{S}$) are selected randomly and compute the gradient of the loss on their local data, then send the compressed weights to the server for a synchronous and update the tier model on server. 

Figure~\ref{system} shows an example of the intra-tier synchronous and cross-tier asynchronous training process. At time $t_1$, the clients in $tier_1$ quickly finish their local training, compress their trained models and send to the server. The server then performs the following steps:
(1) decompresses the local models received from $tier_1$, 
(2) applies synchronous update to the received models of $tier_1$ and get $w^{t_1}_{tier_1}$ (highlighted in red color in Figure~\ref{system}),
and (3) aggregates the latest updates sent from all the tiers (including $tier_1$) using a weighted average aggregation method (see \cref{subsec:weighted_aggregation}), to generate a new global model $w^{t_1}$.

At time $t_2$, the last client of $tier_2$ finishes its local training and sends the compressed model to the server. The server follows the same procedure as $tier_1$ to get a new global model $w^{t_2}$. Then the server sends the latest global model $w^{t_2}$ to the next ready tier (in this example $tier_1$) and starts the next round. Note that a tier in {\model} directly interacts with the server to update the global model whenever it finishes a round of local training, thus forming an \emph{asynchronous}, cross-tier process.

Since clients are partitioned into tiers based on their response latencies and the tiers asynchronously update the global mode, stragglers may not become a performance bottleneck that would otherwise slow down the global training progress. However, as the server interacts more frequently with the faster tiers than with the relatively slower ones, this would inevitably lead to biases towards the faster tiers. To address this issue, we introduce a new objective onto the server-side optimization, which uses a weighted aggregation strategy to more fairly balance the mode updating processes from different tiers.

\subsection{Training at Local Clients}

As mentioned in \cref{preliminaries}, for training with Non-i.i.d. data, frequent local updates may potentially cause the local models to diverge due to the varying updating frequency of different tiers and the underlying data heterogeneity. We add a constraint term to the local subproblem by restricting the local updates to be closer to the global model.
Thus, following \cite{zhang2014deep,li2018federated}, instead of just minimizing the local function $F_k$ , client $k$ applies an update with the constraint using the following surrogate objective $h_k$:
\begin{equation}
    h_k(w_k) = F_k(w_k) + \frac{\lambda}{2}||w_k-w||^2,
\end{equation}
where $w_k$, $w$ are the local model of client $k$ and server model, respectively. 

We use $f_{tier_m}(w)$ as the weighted average of the models from the selected clients within $tier_m$. Assuming at round $t$, $tier_m$ happens to be in communication with the server, we have the update of $f_{tier_m}(w)$ as follows: 
\begin{equation}
\begin{aligned}
&f_{tier_m}(w) = \sum_{k=1}^{|\mathcal{S}_t|}\frac{n_k}{N_c}h_k(w_k) \\
&= \sum_{k=1}^{|\mathcal{S}_t|}\frac{n_k}{N_c}(F_k(w_k) + \frac{\lambda}{2}||w_k-w||^2).
\end{aligned}
\end{equation}
where $\mathcal{S}_t$, $|\mathcal{S}_t|$, and $N_c$ denote a subset of randomly selected clients in $tier_m$,  the number of selected clients in $tier_m$, and the total number of data samples in $\mathcal{S}_t$, respectively. 

The constraint term addresses the issue of Non-i.i.d. by restricting the local updates to be closer to the global model. 
In the ideal situation, with $\lambda=0$, and all clients share the same latency, thus we get one tier and {\model} becomes FedAvg.

\subsection{Cross-Tier Weighted Aggregation}
\label{subsec:weighted_aggregation}

A straightforward idea to achieve unbiased, more balanced training is to assign relatively higher weights to slower tiers that update less frequently, so that the global model would not bias towards the faster tiers.
To this end, {\model} uses a new cross-tier, weighted aggregation heuristic, which dynamically adjusts the relative weights assigned to each tier based on the number of times a tier has updated the global mode. The goal of the weighted aggregation heuristic is to help the global training converge faster. 

Assuming there are $M$ tiers, the number of updates from each tier till now is $T_{tier_1},T_{tier_2},...,T_{tier_M}$, respectively, and the total number of updates from all the tiers till now is $T_{tier_1}+T_{tier_2}+...+T_{tier_M}=T$, we define the objective function of {\model} as: 
\begin{equation}
    f(w) = \sum_{m=1}^{M}\frac{T_{tier_{(M+1-m)}}}{T}f_{tier_m}(w),
\end{equation}
where $\frac{T_{tier_{(M+1-m)}}}{T}$ is the relative weight of $tier_m$, and $\sum_{m=1}^{M}\frac{T_{tier_{(M+1-m)}}}{T}$ $= 1$. To understand the heuristic, a relatively slower tier with a tier number $m$ would get assigned a relatively larger weight value, $\frac{T_{tier_{(M+1-m)}}}{T}$, as $M+1-m$ corresponds to a relatively faster tier, $tier_{(M+1-m)}$, whose historical updating frequency $T_{tier_{(M+1-m)}}$ is expected to be higher. In this way, {\model} can dynamically steer and balance the global model training, avoid potential bias towards a subset of faster tiers, and effectively improve the convergence rate. The approach of {\model} is detailed in Algorithm \ref{alg:FedAT}.
\begin{algorithm}[t]
\caption{{\model}'s Training Process }\label{alg:FedAT}
\KwIn{$w_{tier_m}$, $t$, $T$ and $T_{tier_m}$. $w_{tier_m}$ denotes the weights of Tier $m$. $t$ represents the global round $t$. $T$ is the maximum global rounds. $T_{tier_m}$ is the number of updates of tier $m$}
\textbf{Server:} Initialize $w_{tier_1}, w_{tier_2}...w_{tier_M}$ to $w^{t_0}$. Initialize $t, T_{tier_1}...T_{tier_M}$ to 0\\
\For{each tier m $\in$ $M$ \textbf{in parallel}}{
\While{$t < T$}{
$w^t$ = \textbf{WeightedAverage}($w_{tier_1}, w_{tier_2}...w_{tier_M}$)\\
$\mathcal{S}_m = $ (random set of clients from tier $m$)\\
\For{each client $k$ $\in$ $\mathcal{S}_m$ \textbf{in parallel}}{
$n_k = |\mathcal{D}_k|$\\
$w_k^{t+1} = w_k^t - \eta \nabla h(w^t)$}
$N_c = \sum_{k=1}^{|\mathcal{S}_m|} n_k$\\
$w_{tier_m} = \sum_{k=1}^{|\mathcal{S}_m|} \frac{n_k}{N_c} \cdot w_k^{t+1}$\\
$T_{tier_m} = T_{tier_m} + 1$\\
$t = t + 1$
}
}
\textbf{function WeightedAverage}($w_{tier_1}, w_{tier_2}...w_{tier_M}$)\\
\eIf{$t == 0$}{\textbf{return} $w^{t_0}$}
{\textbf{return} $\sum_{m=1}^{M} \frac{T_{tier_{(M+1-m)}}}{T} \cdot w_{tier_m}$}

\end{algorithm}

\subsection{Compression, Marshalling, and Unmarshalling}
\label{compress-sec}

Previous work on communication-efficient FL mentioned in \cref{sec:related} almost exclusively consider i.i.d. data distribution among the clients, which is not practical in
real-world scenarios of FL, where the client data typically follows a Non-i.i.d. distribution~\cite{mcmahan2017communication}.
As studied in~\cite{sattler2019robust}, many compression methods~\cite{bernstein2018signsgd,sattler2019sparse} suffer from slow convergence rates in the Non-i.i.d. cases. Non-i.i.d. often introduces divergence of model weights collected from resource-heterogeneous clients. Due to such divergence, some lossy compression methods such as quantization and dequantization~\cite{batchcrypt_atc20} may inevitably lead to huge errors and reduce global performance. Furthermore,
as asynchronous FL methods aggregate more frequently than synchronous FL methods, highly frequent updates drastically amplify such divergence, and result in a poor global performance. Therefore, with frequent communications in asynchronous FL approaches, it is crucial to select a compression technique that can efficiently reduce the communication cost while effectively guaranteeing a fast convergence to the optimal solution.

To this end, we design a simple yet effective compression scheme based on Encoded Polyline Algorithm \footnote{\url{https://developers.google.com/maps/documentation/utilities/polylinealgorithm}}  (or polyline encoding).
Polyline encoding is a lossy compression algorithm that converts a rounded binary value into a series of character codes for ASCII characters using the base64 encoding scheme.
It can be configured to maintain a configurable precision by rounding the value to a specified decimal place. With the appropriate precision, the model could achieve the largest communication saving and minor performance loss.
Our experiments show that it can achieve a high compression ratio of up to $3.5\times$ under the FL communication scenarios. 
(We evaluate the effectiveness of compression in \cref{subsec:compression_eval}.) 
{\model} compresses both the uplink and downlink communications. 
The process is as follow:
(1) {\model} flattens (marshalling) the weights of each layer to get a list of decimal values. 
(2) Then, using polyline encoding, every decimal value in the list gets converted into a compressed ASCII string; 
along with the compressed weights, the dimensions of the weights of each layer is transmitted as well. 
(3) When the server/clients receive the compressed weights, a decompression process is performed, and then the decompressed weights are reshaped back (unmarshalling) to the original dimensions based on the dimension information received.\\

\section{Convergence Analysis}

In this section, we show that \model~converges to the global optimal solution for both strongly convex and non-convex functions on Non-i.i.d. data in theory. It is consistent with our experiments results shown in Figure~\ref{non-iid}.
First, we introduce the definitions and assumptions as follows. 
\begin{definition}\label{smoothness}
(Smoothness) The function $f$ has Lipschitz continuous gradients with constant $L > 0$ (in other words, $f$ is \textit{L-smooth}) if $\forall x_1,x_2$,
\begin{equation}
f(x_1)-f(x_2) \leq \langle\nabla f(x_2), x_1 - x_2\rangle + \frac{L}{2}||x_1 - x_2||^2.
\end{equation}
\end{definition}
\begin{definition}\label{strong_convex}
(Strong convexity) The function $f$ is $\mu$-\textit{strongly convex} with $\mu > 0$ if $\forall x_1,x_2$,
\begin{equation}
f(x_1)-f(x_2) \geq \langle\nabla f(x_2), x_1 - x_2\rangle + \frac{\mu}{2}||x_1 - x_2||^2.
\end{equation}
\end{definition}

Definition \ref{inexactness} has been made by the work \cite{li2018federated}.

\begin{definition}\label{inexactness}
($\gamma$-inexactness) For a function $h(w) = F(w) + \frac{\lambda}{2}||w-w_0||^2$, and $\gamma \in [0,1]$. $w^*$ is a $\gamma$-inexact solution for $\min_w h(w)$ if $||\nabla h(w^*)|| \leq \gamma||\nabla h(w_0)||$, where $\nabla h(w) = \nabla F(w) + \lambda(w-w_0)$.  
\end{definition}

According to \cite{li2020federated}, this definition aims to allow flexible performance of local clients in each communication round, such that each of
the local objectives can be solved inexactly. The amount of local computation vs. communication can be tuned
by adjusting the number of local iterations, i.e., more local iterations indicates more exact local solutions.

Further, we make the following assumptions on the objective functions:

\begin{assumption}\label{bound_min}
The central objective $f(w)$ is bounded, i.e., $\min f(w) = f(w_*) > -\infty$.
\end{assumption}

\begin{assumption}\label{bound_gradient}
The expected squared norm of stochastic gradients is uniformly bounded, i.e., there exists a scalar $G$, such that $\mathbb{E}||\nabla F(w^t)||^2 \leq G^2$, all $t = 0,...,T-1$. 
\end{assumption}

\begin{assumption}\label{bound_gradient2}
 With $\bar g_t(w^t)$ ($\bar g_t = \sum_{k=1}^{m} \frac{n_k}{N_c} \nabla h_k(w^t)$) as the averaged gradients from a certain tier with $m$ clients, there exists a scalar $\sigma > 0$ such that $\nabla f(w^t)^{\top} \mathbb{E}(\bar g_t(w^t)) \geq \sigma ||\nabla f(w^t)||^2 $.  
\end{assumption}
Assumption \ref{bound_min} is easy to satisfy as there exists a minimum value for the central objective $f(w)$. The conditions in Assumption \ref{bound_gradient} on the variance of stochastic gradients is customary. While this is a much weaker assumption compared to the one that uniformly bounds the expected norm of the stochastic gradient. Assumption \ref{bound_gradient2} ensure that the gradient of local tier $\bar g_t$ is an estimation of $\nabla f(w^t)$. And as $\sigma = 1$, we have $\bar g_t$ as the unbiased estimation of $\nabla f(w^t)$. 

To convey our proof clearly, we first introduce and prove certain useful lemmas.
\begin{lemma}\label{bound_g}
With Definition \ref{inexactness}, the local functions $h(\cdot)$ are $\gamma$-inexact. For aggregated tier model $\bar g_t(w^t)$, we have
\begin{equation}
    \mathbb{E}||\bar g_t(w^t)||^2 \leq \gamma^2 G^2 c^2,
\end{equation}
where $c$ is the total number of clients within the given tier. 
\end{lemma}

\begin{lemma}
\label{lemma_2}
If $f(w)$ is $\mu$-\textit{strongly convex}, then with Definition  \ref{strong_convex}, we have: 
\begin{equation}
2\mu(f(w^t) - f(w_*)) \leq  ||\nabla f(w^t)||^2.
\end{equation}
\end{lemma}

We show that the averaged model of local tier is bounded in Lemma \ref{bound_g}. The detailed proof is in Appendix \ref{lemma1_proof}. While the proof of Lemma \ref{lemma_2} is supported by the literature \cite{nesterov2013introductory,bottou2018optimization}, we also provide a detailed proof in Appendix \ref{lemma2_proof}.

\begin{theorem}\label{convex_theo}
Suppose that the central objective function $f(w)$ is \textit{L-smooth} and $\mu$-\textit{strongly convex}. The local functions $h(\cdot)$ are $\gamma$-inexact. Let \textit{Assumption} \ref{bound_gradient} and \textit{Assumption} \ref{bound_gradient2} hold.  
After $T$ global
updates on the server, \model~converges to a global optimum $w_*$: 
\begin{equation}\label{convex_con}
\begin{aligned}
    &\mathbb{E}[f(w^T) - f(w_*)] \\
    &= (1-2\mu B \eta \sigma)^T(f(w^0)-f(w_*))+ \frac{L}{2} \eta^2 \gamma^2 B^2 G^2 c^2, 
\end{aligned}
\end{equation}
\end{theorem}
where $c$ is the total number of clients within one tier, and $B = \frac{T_{tier_{(M+1-m)}}}{T} \leq 1$. 

We direct the reader to Appendix \ref{convex_proof} for a detailed proof.  The convergence bound in Theorem \ref{convex_theo} depends on the local constrain $\mu$, weighted parameter $B$ and learning rate $\eta$. Note that $B$ varies in each global iteration because the update number of each tier changes at every global iteration. 

\begin{theorem}\label{nonconvex_theo}
Suppose that the central objective function $f(w)$ is \textit{L-smooth} and \textit{non-convex}. The local functions $h(\cdot)$ are $\gamma$-inexact. Let \textit{Assumption} \ref{bound_min}, \textit{Assumption} \ref{bound_gradient} and \textit{Assumption} \ref{bound_gradient2} hold,  
then after $T$ global
updates we have:
\begin{equation}\label{nonconvex_con}
\begin{aligned}
    &\sum_{t=0}^{T-1} B \mathbb{E}[||\nabla f(w_t)||^2] \\ 
    &\leq \frac{f(w^0) -f(w_*)}{\eta \sigma} + \frac{L}{2\sigma} T^2\eta \gamma^2 B G^2 c^2, 
\end{aligned}
\end{equation}
\end{theorem}
where $c$ is the total number of clients within one tier, and $B = \frac{T_{tier_{(M+1-m)}}}{T} \leq 1$.

\section{Experimental Setting}
\label{subsec:methodology}

\noindent\textbf{Federated Datasets.} 
We evaluate {\model} using five different federated datasets and an FL benchmarking framework \textbf{LEAF}~\cite{caldas2018leaf} on both convex and non-convex models described as follows:
\begin{itemize}[noitemsep,leftmargin=*]
    \item{CIFAR-10:}  The CIFAR-10~\cite{krizhevsky2009learning} dataset consists of $60,000$ $32\times32$ colour images in $10$ classes, with $6000$ images per class. There are $50,000$ training images and $10,000$ test images. We partition the dataset into $100$ clients and follow the same Non-i.i.d. setting of CIFAR-10 as \cite{mcmahan2017communication}. 
    
    \item{Fashion-MNIST:} Fashion-MNIST~\cite{xiao2017fashion} is a dataset that contains a training set of $60,000$ examples and a test set of $10,000$ examples. Each example is a $28\times28$ grayscale image, associated with a label from $10$ classes. We use the same number of clients and Non-i.i.d. setting as CIFAR-10.  
    
    \item{Sentiment140:} Sentiment140~\cite{go2009twitter} is a text dataset that contains $1,600,000$ tweets, and the tweets are labeled with one of negative, neutral and positive. Each twitter account corresponds to a client.
    
    \item{FEMNIST:} FEMNIST~\cite{caldas2018leaf} is an image dataset that consists of $805,263$ samples. There are a total of $62$ classes, where all the samples are distributed to $3,550$ clients with inherent data heterogeneity and Non-i.i.d.-ness. 

    \item{Reddit:}
    Reddit~\cite{caldas2018leaf} is a text dataset that contains processed comments posted on Reddit in December 2017. It has a total of $56,587,343$ samples collected from $1,660,820$ Reddit users.
    
\end{itemize}

\noindent\textbf{FL Methods.}
We compare {\model} against five synchronous and asynchronous FL methods:
\begin{itemize}[noitemsep,leftmargin=*]
    \item{FedAvg~\cite{mcmahan2017communication}:} A baseline synchronous FL method proposed by McMahan et al. At each round, a certain ratio of total clients are randomly selected for training, the server aggregates the weights received from selected clients in an average manner.
    \item{TiFL~\cite{chai2020tifl}:} A synchronous FL method that partitions training clients into different tiers based on their responding latency. For each round, one tier is selected according to a novel adaptive selection strategy which is related to the test accuracies of all tiers, then certain number of clients in that tier are selected for training. The aggregation method of TiFL is adopted from FedAvg. 
    \item{FedProx~\cite{li2018federated}:} A synchronous FL method that claims to handle stragglers due to system heterogeneity across all clients by applying different local epochs for clients. Also FedProx adds proximal term to the objective on the clients for improving performance and smoothing the training curve.
    
    \item{FedAsync~\cite{xie2019asynchronous}:} A baseline asynchronous FL method using weighted averaging to update the server model. Different from previous synchronous FL methods, all clients train concurrently, when the server receives weights from any client, the weights are weighted averaged with current global weights to get the latest global weights, then communicate to all available clients at that time for training.
    
    \item{ASO-Fed~\cite{chen2019asynchronous}:} An asynchronous FL framework designed for online data. Unlike FedAsync, ASO-Fed adds local constraints on the client side. ASO-Fed maintains a copy of weights for each client on the server side, and it calculates global weights by averaging all the copies. Note that we compare ASO-Fed against {\model} on a large-scale, 500-client cluster on FEMNIST (\cref{subsec:large_scale}).
\end{itemize}

\begin{table*}[h]
\centering
\caption{Comparison of prediction performance and variance to baseline approaches. {\small{\texttt{\#class}}} indicates the number of labels (i.e., classes) each client has. 
The {\small{\texttt{Accuracy}}} rows show the best prediction accuracy that each FL approach reaches after each model converges. 
The {\small\texttt{Norm.Var.}} rows show the average variance of test accuracy among all clients, normalized to that of {\model}. 
We show {\model}'s absolute variance values ({\small\texttt{Abs.Var.}}).
We show the absolute values for {\model}'s accuracy variance. 
{\small\texttt{impr.(a)}} and {\small\texttt{impr.(b)}} are the accuracy improvement of {\model} compared with the best and worst baseline FL method, respectively. 
The best performance results are highlighted in bold font. 
} 
\label{table:performance}
\vspace{-10pt}
\resizebox{.8\textwidth}{!}{\begin{tabular}{llccccccc}
\hline
\multicolumn{2}{c}{\multirow{2}{*}{Dataset(\#class)}}                & \multicolumn{5}{c}{CIFAR-10}                                                            & Fashion-MNIST  & \multirow{2}{*}{Sentiment140}  \\ \cmidrule(r){3-7}\cmidrule(r){8-8}
\multicolumn{2}{c}{}                                                 & \#2             & \#4             & \#6             & \#8            & i.i.d.           & \#2            &                                \\ \hline
\multicolumn{1}{l|}{\multirow{2}{*}{\textbf{TiFL}}}     & Accuracy   & 0.527           & 0.615           & 0.654           & 0.655          & 0.685            & 0.859          & 0.739                          \\
\multicolumn{1}{l|}{}                                   & Norm. Var. & 1.26            & 2.79            & 1.33            & 1.3            & 2.12             & 1.29           & 2.75                           \\ \hline
\multicolumn{1}{l|}{\multirow{2}{*}{\textbf{FedAvg}}}   & Accuracy   & 0.547           & 0.628           & 0.654           & 0.667          & 0.686            & 0.842          & 0.741                          \\
\multicolumn{1}{l|}{}                                   & Norm. Var. & 2               & 5.07            & 4.33            & 3.1            & 4.23             & 1.86           & 3.72                           \\ \hline
\multicolumn{1}{l|}{\multirow{2}{*}{\textbf{FedProx}}}  & Accuracy   & 0.509               & 0.609               & 0.624               & 0.650              & 0.669                & 0.831              & 0.742                              \\
\multicolumn{1}{l|}{}                                   & Norm. Var. & 1.261               & 6.75               & 3.981               & 2.22              & 2.992                & 2.243             & 3.89                              \\ \hline
\multicolumn{1}{l|}{\multirow{2}{*}{\textbf{FedAsync}}} & Accuracy   & 0.480            & 0.541           & 0.531           & 0.561          & 0.567            & 0.795          & 0.740                          \\
\multicolumn{1}{l|}{}                                   & Norm. Var. & 2               & 3.93            & 2.08            & 1.54           & 2.69             & 2              & 5.69                           \\ \hline
\multicolumn{1}{l|}{\multirow{4}{*}{\textbf{FedAT}}}    & Accuracy   & \textbf{0.591}  & \textbf{0.633}  & \textbf{0.673}  & \textbf{0.681} & \textbf{0.701}   & \textbf{0.873} & \textbf{0.748}                 \\
\multicolumn{1}{l|}{}                                   & Abs. Var.  & \textbf{0.0042} & \textbf{0.0014} & \textbf{0.0012} & \textbf{0.001} & \textbf{0.00052} & \textbf{0.007} & \textbf{$\mathbf{2.67e^{-5}}$} \\
\multicolumn{1}{l|}{}                                   & impr.(a)   & 7.44\%          & 0.79\%          & 2.82\%          & 2.05\%         & 2.13\%           & 1.6\%          & 0.93\%                         \\
\multicolumn{1}{l|}{}                                   & impr.(b)   & 18.78\%         & 14.53\%         & 21.09\%         & 17.62\%        & 19.11\%          & 8.93\%         & 1.2\%                          \\ \hline
\end{tabular}}
\vspace{-10pt}
\end{table*}

\begin{figure*}[h]
    \centering
    \includegraphics[scale =0.35,trim=30 85 10 50,clip]{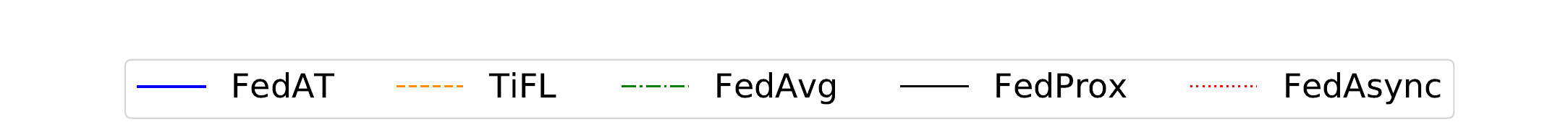}
    \label{fig:full_legend}
\vspace{-1em}
\end{figure*}
\begin{figure*}[h]
  \centering
  \subfigure[CNN $@$ CIFAR-10. ]{\includegraphics[scale=0.33,trim=0 0 0 5, clip]{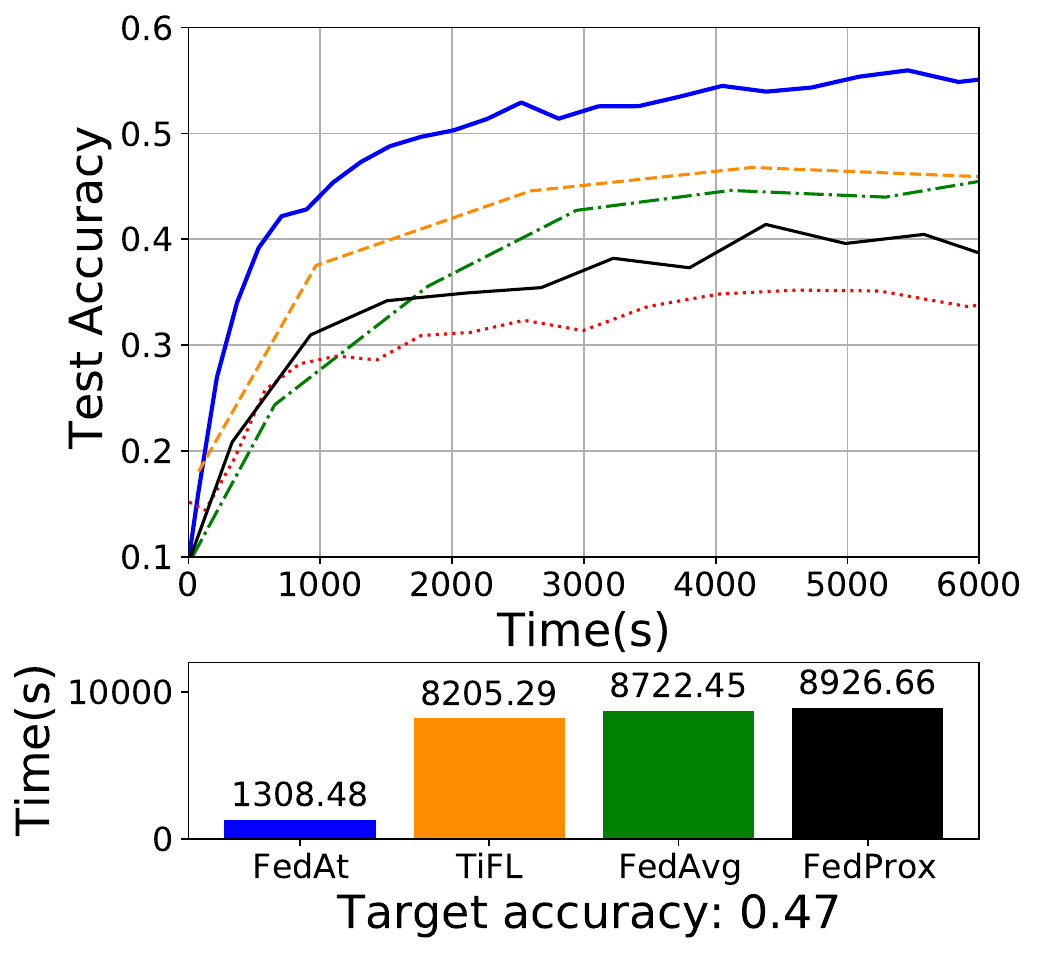}}\label{fig:converge-all-a}
  \subfigure[CNN $@$ Fashion-MNIST.]{\includegraphics[scale=0.33,trim=0 0 0 5, clip]{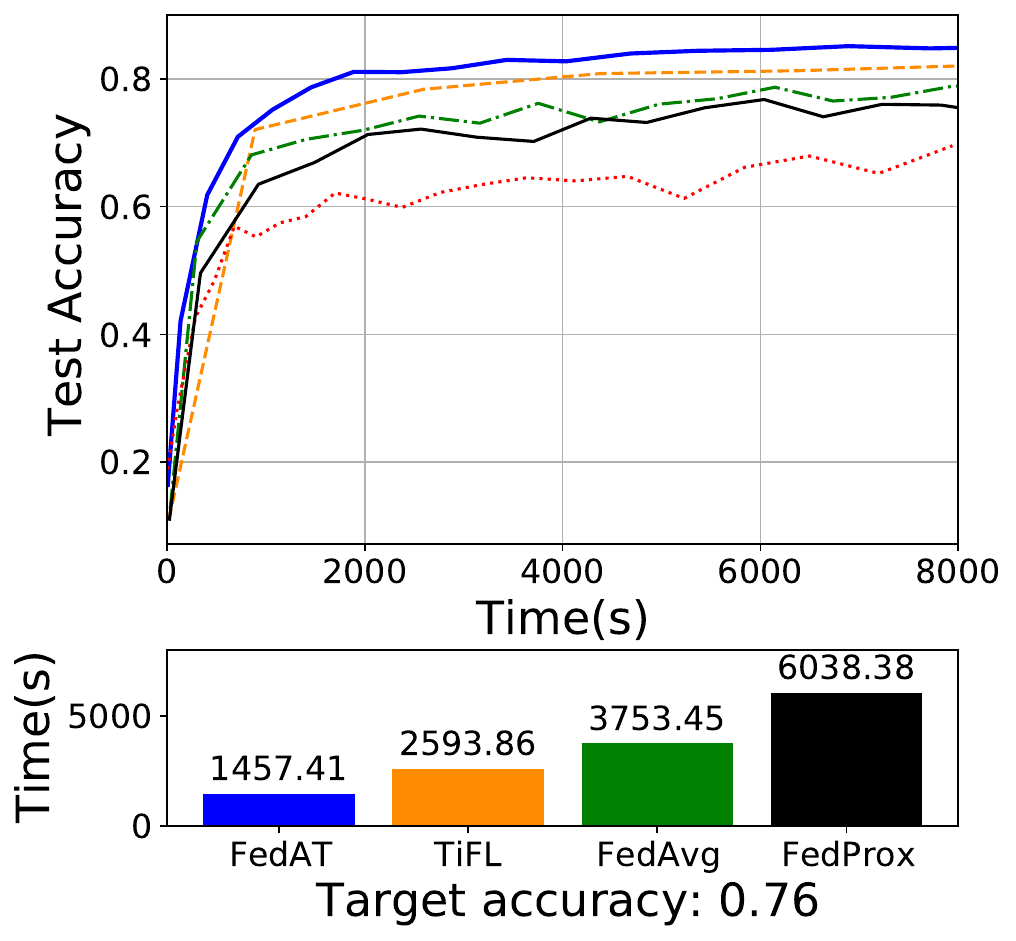}}
  \subfigure[Logistic $@$ Sentiment140.]{\includegraphics[scale=0.33,trim=0 0 0 5, clip]{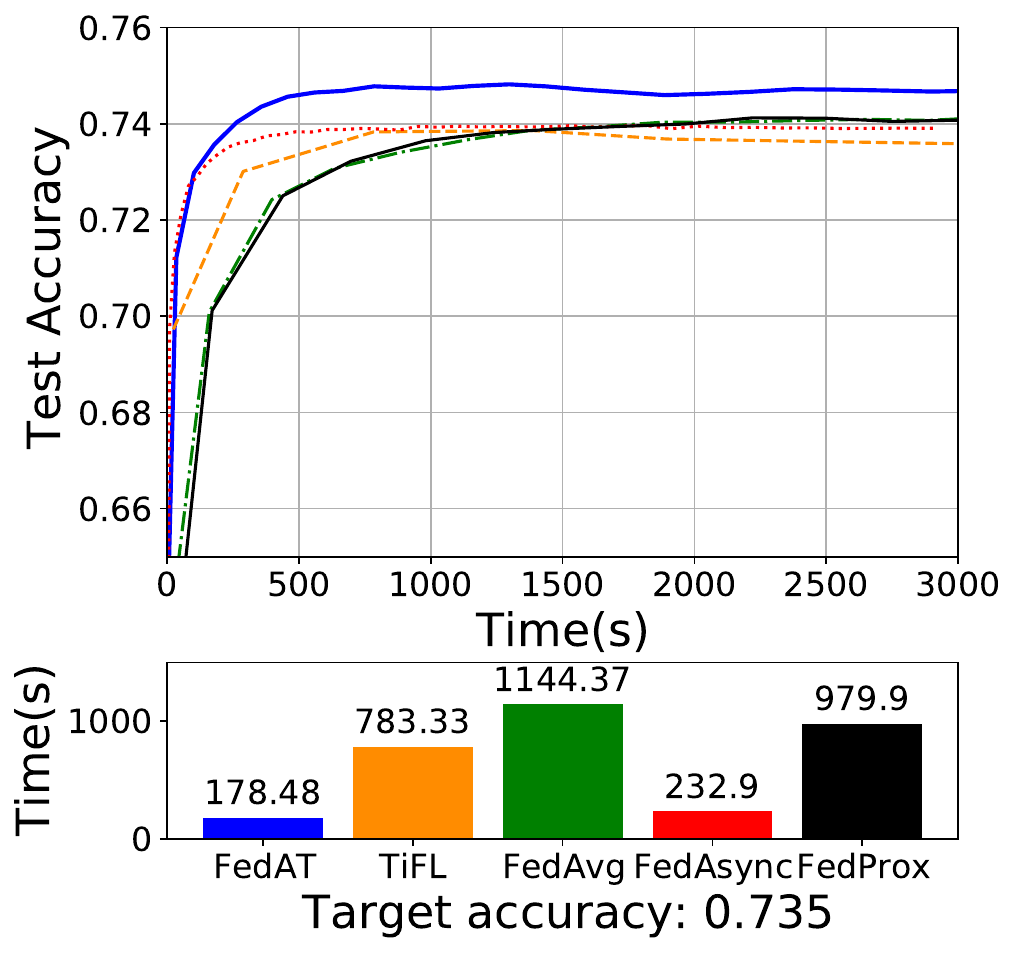}}
  \caption{
  Performance comparison of different FL methods on 2-class Non-i.i.d. CIFAR-10, Fashion-MNIST, and Sentiment140 datasets. 
  \textbf{Above (test accuracy timeline curves):} The results are average-smoothed for every 40 global rounds. 
  \textbf{Bottom (bar charts):} the time it takes for each evaluated FL methods to reach a target accuracy of $N\%$ as specified in the X-axis' labels. (Note that FedAsync is not able to reach the target accuracy for CIFAR-10 and Fashion-MNIST, thus is omitted.)
  }\label{converge-all}
\end{figure*}

\section{Evaluation}
\label{sec:eval}

\noindent\textbf{Implementation and Setup.}
We have implemented {\model} and the comparison FL methods all in TensorFlow~\cite{abadi2016tensorflow}. 
For all the experiments except FEMNIST and Reddit, we simulate a FL setup using a 192-core cluster on Chameleon Cloud~\cite{chameleon_cloud, chameleon_atc20}, which consists of three bare-metal servers, each with a 64-core Intel$^\circledR$ Xeon$^\circledR$ Gold 6242 CPU, and 192 GB main memory; 
we deploy the FL server exclusively on one server, and all clients on the other two servers, where each client gets assigned one CPU core. We evaluate 100 clients in Chameleon Cloud tests. 

For experiments on FEMNIST and Reddit, we deploy the FL server exclusively on one \textit{c5.24xlarge} virtual machine (VM) instance (96 vCPUs and 192~GB memory) on AWS and $500$ participating clients on a large-scale, $100$-VM cluster, where each VM is a
\textit{c5.2xlarge} instance with 8 vCPUs and 16~GB memory. 

{\model} is configured to use {\small\texttt{Precision 4}} as the precision of the compressor (\cref{subsubsec:compression}) throughout the evaluation for CIFAR-10, Fashion-MNIST and Sentiment140.

\noindent\textbf{Models.}
We train a convolutional neural network (CNN) on CIFAR-10 and Fashion-MNIST. The network architecture includes three convolutional layers, each with $32$, $64$ and $64$ filters, followed by two fully connected layers with units of $64$ and $10$. For Sentiment140, we train a logistic regression model to evaluate the model performance under a convex setting. 
For the FEMNIST dataset, we train a similar CNN that classifies images. For the Reddit dataset, we train an LSTM model~\cite{lstm}. The LSTM model starts with an embedding layer with an input dimension of $10,000$ and an output dimension of $128$, followed by an LSTM layer with a dropout rate of 0.1, a batch normalization layer and a dense layer with $10,000$ units.

\noindent\textbf{Hyperparameters.}
We randomly split each client's local data into an 80\% training set and a 20\% testing set. For intra-tier synchronous training, we adopt the same sampling scheme as FedAvg: sampling clients (within a particular tier) randomly at each round. We use Adam~\cite{kingma2014adam} as the local solver and set the local constrain parameter $\lambda$ as $0.4$. For each dataset, we tune the learning rate for FedAvg using the following configuration: \textit{local epoch} $E=3$, \textit{batch size} $= 10$; 
we use the same learning rate for the other five FL methods.
We set the number of randomly selected clients as $10$ for FedAvg, TiFL, FedProx, and {\model} on all datasets.

\noindent\textbf{Simulating Different Performance Tiers.}
Clients in FL are typically edge devices, and their computing power and network connection may not be stable; hence simply assigning a fixed amount of resources is not sufficient to reflect the real situation.
Therefore, we assign 1 CPU for each client during the whole training process and add random delays to the computations conducted by clients; the added random delay is to simulate different levels of straggler effects that are caused by weaker computing powers and intermittent network connections in a real-world FL setup.
We first evenly divide all the clients into $5$ parts, then randomly assign delays of $0s$, $0 \sim 5s$, $6 \sim 10s$, $11 \sim 15s$, and $20 \sim 30s$ to the clients in each part at every round, respectively. Each part is called one \emph{tier}. 
To guarantee fair comparison,
each client, once selected, would follow a fixed, pseudo-random mini-batch schedule. The same strategy is applied to all the FL methods that we test (including {\model}'s intra-tier synchronous training). Furthermore, to simulate unstable network connections, for all the tests that we run, we randomly select 10 ``unstable'' clients, which would drop out at any time during the training process. Once the client drops out, it will not come back and rejoin the training process again.

\subsection{Prediction Performance}

Table~\ref{table:performance} presents the results of the prediction performance and the variance of the test accuracy on all the datasets. We report the best test accuracy after each training process converges within a global iteration budget. For the 2-class CIFAR-10 dataset, {\model} outperforms the best baseline FL method, FedAvg, by $7.44\%$, and the worst baseline method, FedAsync, by $18.78\%$. 
Using the same tiering scheme as TiFL, {\model} achieves consistently higher accuracy than TiFL for all the experiments. 
This is because: 
(1) the local constraint forces local models to be closer to the server model,
and (2) {\model}'s new weighted aggregation heuristic can more effectively engage the straggling clients from the slower tiers, leading to better prediction performance (we evaluate the effectiveness of our weighted aggregation method in \cref{com-aggregation}). 
FedAvg has the closest prediction performance as TiFL, because they both follow the same synchronous updating strategy. FedAsync, on the other hand, performs the worst, 
as it simply aggregates weights from one client at a round and has no effective way to deal with stragglers.
The performance difference can also be clearly noticed from the convergence timeline graphs shown in Figure~\ref{converge-all}. {\model} converges faster towards the optimal solution than all other three compared methods on both the non-convex and convex objectives. 
\begin{figure*}[t]
    \centering
    \includegraphics[scale =0.35,trim=30 85 10 50,clip]{charts/full_legend.pdf}
    \label{fig:full_legend}
\vspace{-1em}
\end{figure*}
\begin{figure*}[t]
  \centering
  \subfigure[CIFAR-10 Non-i.i.d. (4). ]{\includegraphics[scale=0.27]{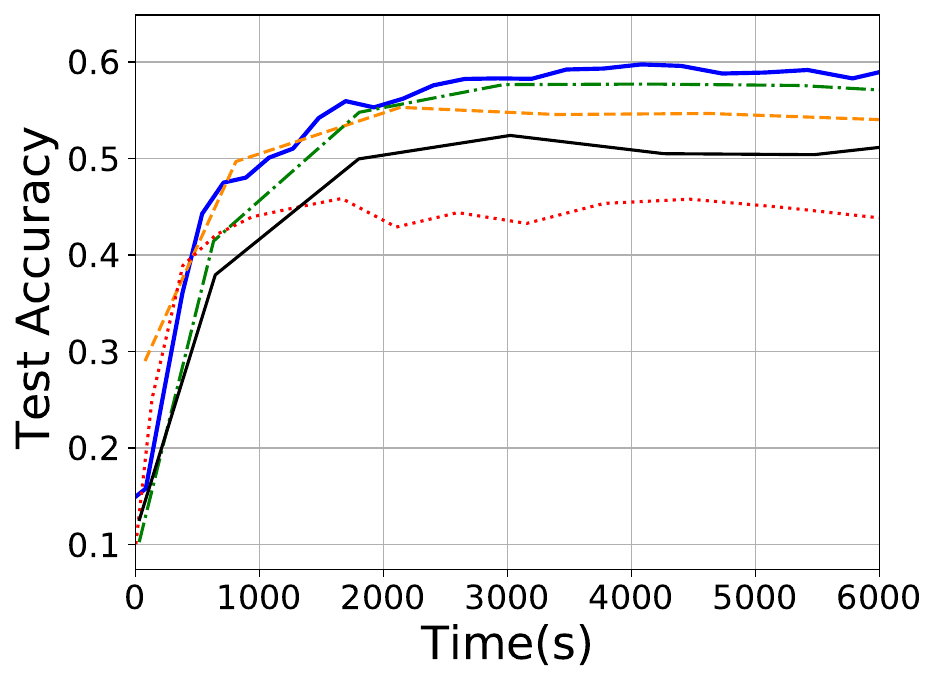}}
  \subfigure[CIFAR-10 Non-i.i.d. (6).]{\includegraphics[scale=0.27]{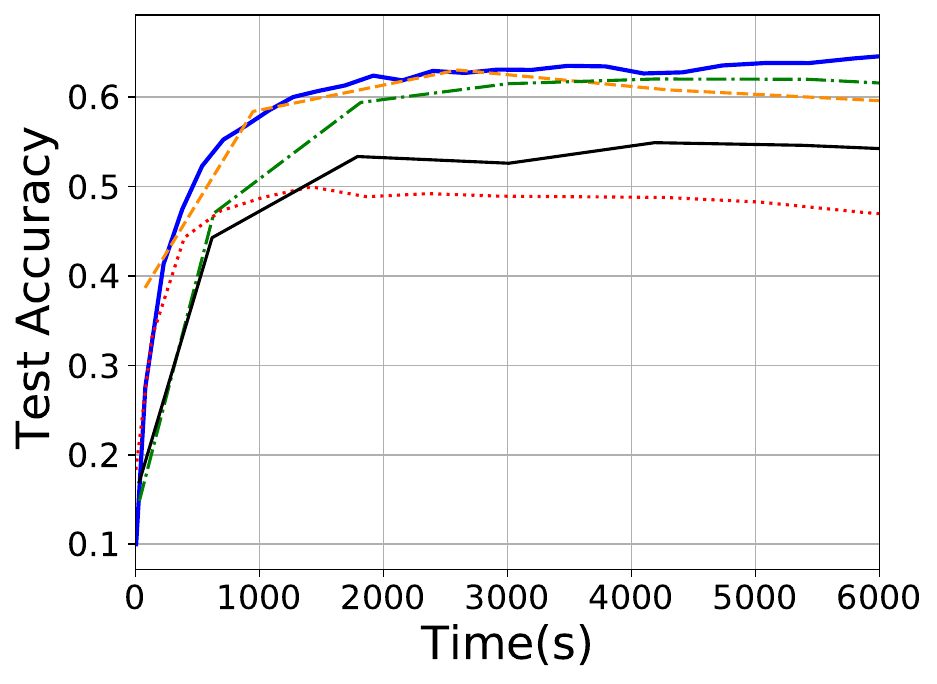}}
  \subfigure[CIFAR-10 Non-i.i.d. (8).]{\includegraphics[scale=0.27]{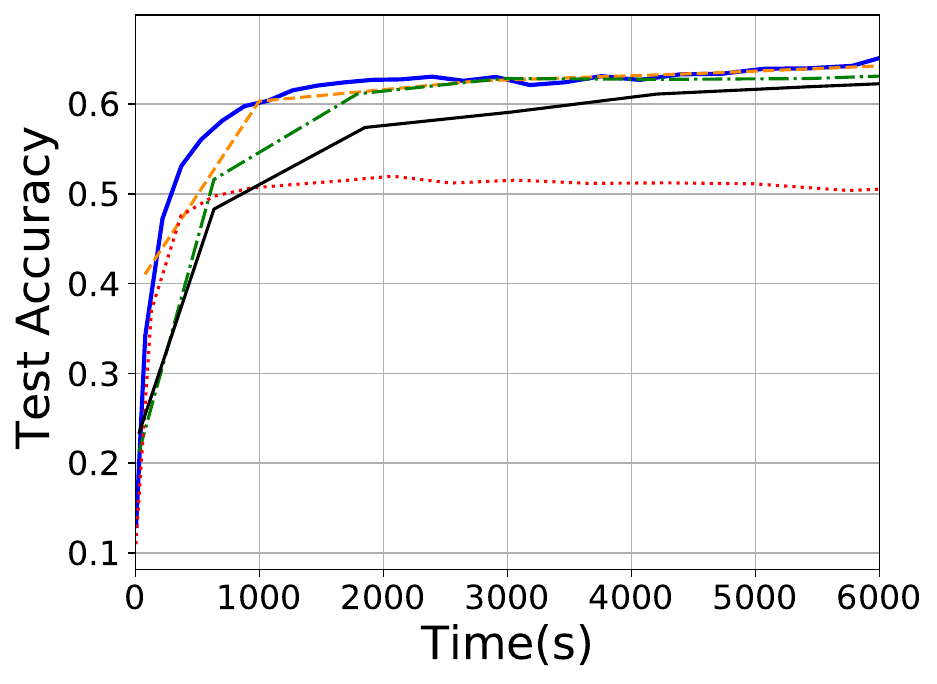}}
  \subfigure[CIFAR-10 i.i.d.]{\includegraphics[scale=0.27]{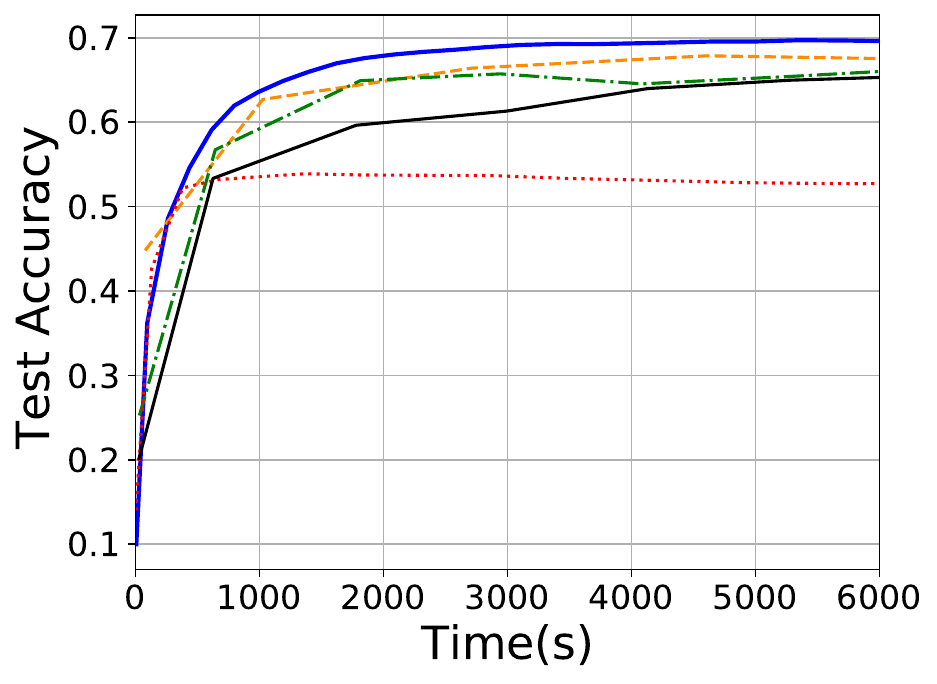}}
\vspace{-1em}
  \caption{
  Convergence speed comparison on CIFAR-10 over different level of Non-i.i.d.-ness. The results are average-smoothed for every 40 global rounds. 
  }\label{non-iid}
  \vspace{-1.5em}
\end{figure*}

\begin{table}[h]
\centering
\caption{
Amounts of data (MB) transferred between clients and server to achieve the target accuracy on all datasets with 2-class Non-i.i.d. case. {\small\texttt{--}} means that the FL method is not able to achieve the accuracy target within the iteration budget. The best results are highlighted in bold font. 
}\label{compress-byte}
\vspace{-5pt}
\resizebox{0.48\textwidth}{!}{\begin{tabular}{l|c|l|c|l|c}
\hline
\multirow{2}{*}{Method} & \multicolumn{2}{c|}{CIFAR-10}          & \multicolumn{2}{c|}{Fashion-MNIST}           & Sentiment140   \\
                        & \multicolumn{2}{c|}{(acc. = 0.50)}    & \multicolumn{2}{c|}{(acc. = 0.79)}    & (acc. = 0.73)  \\ \hline
FedAvg                  & \multicolumn{2}{c|}{1828.54}          & \multicolumn{2}{c|}{1048.25}          & 16.71          \\ \hline
TiFL                    & \multicolumn{2}{c|}{2140.71}          & \multicolumn{2}{c|}{1041.98}          & 17.20          \\ \hline
FedProx                 & \multicolumn{2}{c|}{$-$}              & \multicolumn{2}{c|}{2169.95}          & 18.42          \\ \hline
FedAsync                & \multicolumn{2}{c|}{$-$}              & \multicolumn{2}{c|}{9895.53}          & 82.27          \\ \hline
FedAT                   & \multicolumn{2}{c|}{\textbf{1675.82}} & \multicolumn{2}{c|}{\textbf{1041.54}} & \textbf{16.41} \\ \hline
\end{tabular}}
\end{table}

\begin{figure*}[h]
    \centering
    \includegraphics[scale =0.3,trim=30 85 10 50,clip]{charts/full_legend.pdf}
    \label{fig:full_legend}
\vspace{-1em}
\end{figure*}
\begin{figure*}[h]
  \centering
  
  \subfigure[CIFAR-10.]{\includegraphics[scale=0.3]{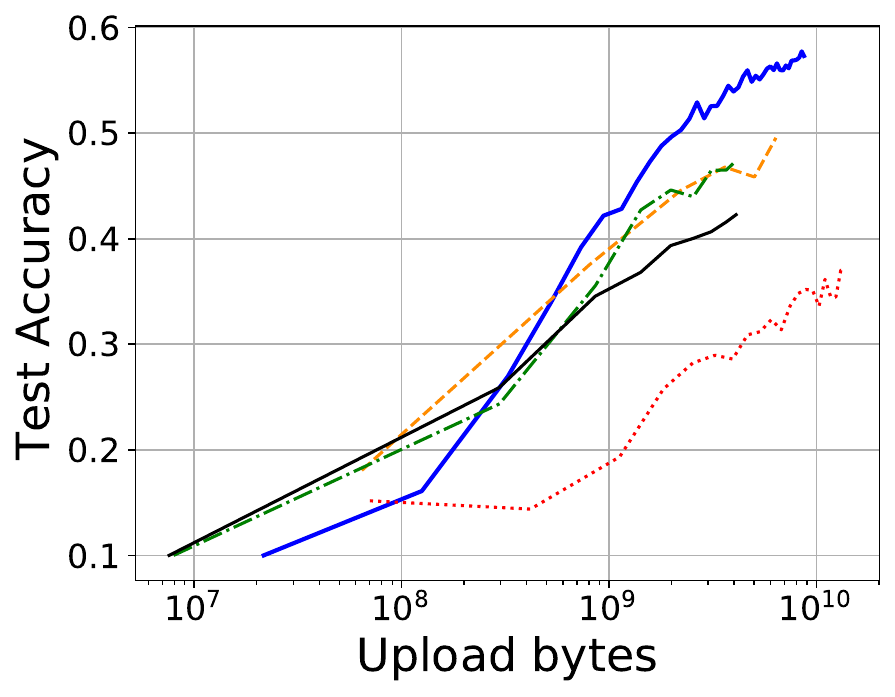}}
  \subfigure[Fashion-MNIST.]{\includegraphics[scale=0.3]{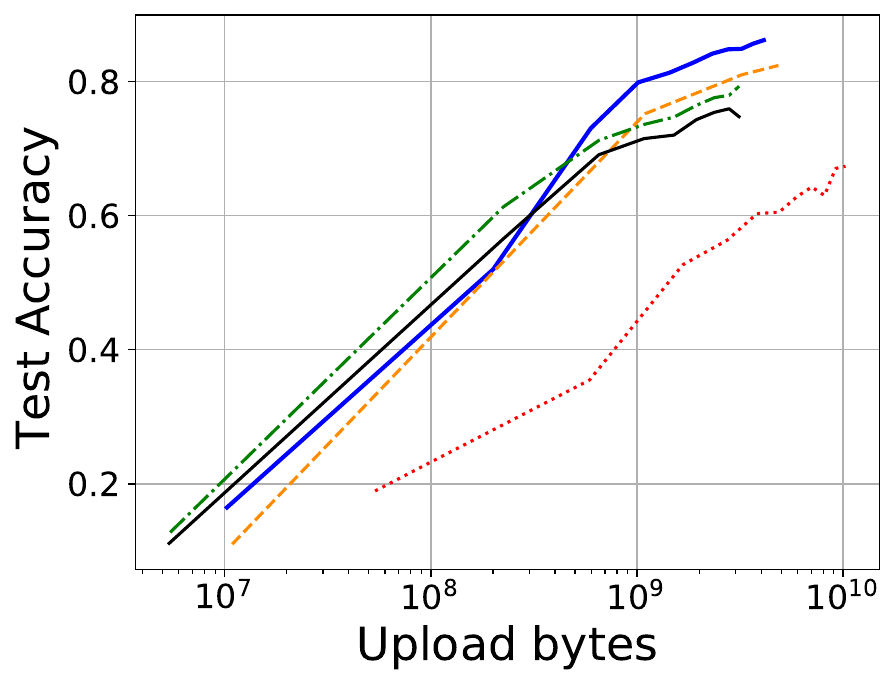}}
  \subfigure[Sentiment140.]{\includegraphics[scale=0.3]{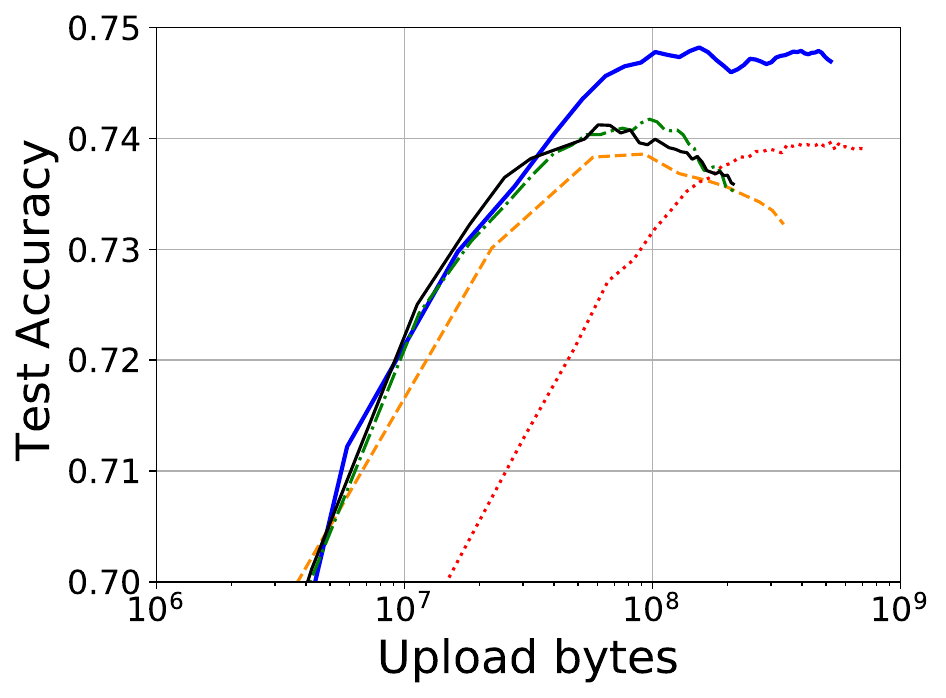}}
      \vspace{-1em}

  \caption{ Test accuracy as a function of the cumulative amounts of data uploaded from clients to the server for 2-class Non-i.i.d. datasets. The performance curves are average-smoothed for every $40$ global rounds. The X-axis is in log-scale. 
  }\label{compress-curve}
    \vspace{-5pt}

\end{figure*}

\vspace{-6pt}
\subsubsection{Impact of Non-i.i.d. Level.}

The models' convergence behaviors are sensitive to the degree of Non-i.i.d. of the data distribution across clients.  Table~\ref{table:performance} shows that, for the CIFAR-10 dataset, the test accuracy increases as the degree of Non-i.i.d. decreases (i.e., the number of classes per client increases); accordingly, the variance of the test accuracy decreases as the degree of Non-i.i.d. decreases (i.e., the data is more evenly shuffled and each client covers all the classes). Figure~\ref{converge-all}(a) (the timeline charts above) and \ref{non-iid} together show a sensitivity analysis of the convergence rate as a function of the Non-i.i.d. (from two classes per client, to 8 classes per client, to the i.i.d. case), on the CIFAR-10 dataset. We observe that {\model} outperforms all the other four FL methods with higher prediction performance across all different Non-i.i.d. levels. The most distinct performance gap between {\model} and the other FL methods can be observed in the 2-class Non-i.i.d. case, where each individual client holds only $2$ classes of data. Notably, {\model} improves the prediction performance by as much as $8.04\%$ compared to FedAvg.

\subsubsection{Robustness to Stragglers.}

As defined in in Definition~\ref{defi-straggle}, the robustness of a FL method against stragglers can be quantified using the variance on the prediction performance and the convergence speed. Table~\ref{table:performance} shows that {\model} has consistently the lowest accuracy variance across all experiments. FedAvg observes significantly higher accuracy variance, which are 1.86-5.07$\times$ higher than that of {\model}. This is due to the compound effect of both synchronous training and stragglers -- synchronous training determines that during each round only a subset of clients can get involved to contribute to the global training, while the straggling clients are more likely to have a less accurate model when they next get selected (since they receive less training) by the server for training, thus causing huge accuracy fluctuation of the global model.

The bar charts in Figure~\ref{converge-all} presents a comparison of the training time it takes for each FL method to achieve a target test accuracy. For example, as shown in Figure~\ref{converge-all}(a) (bar chart at bottom), to reach an accuracy of $47\%$ for the CIFAR-10 CNN model, TiFL, FedAvg and FedProx spend $5.27\times$, $5.67\times$ and $5.82 \times$ longer time than {\model}. Fashion-MNIST show a similar trend. For Sentiment140, 
TiFL, FedAvg, FedProx and FedAsync take $3.39\times$, $5.41\times$, $4.49\times$ and $0.3\times$ longer time than {\model}, respectively.

\begin{figure}[t]
  \centering
  \subfigure[Test accuracy. ]{\includegraphics[scale=0.27]{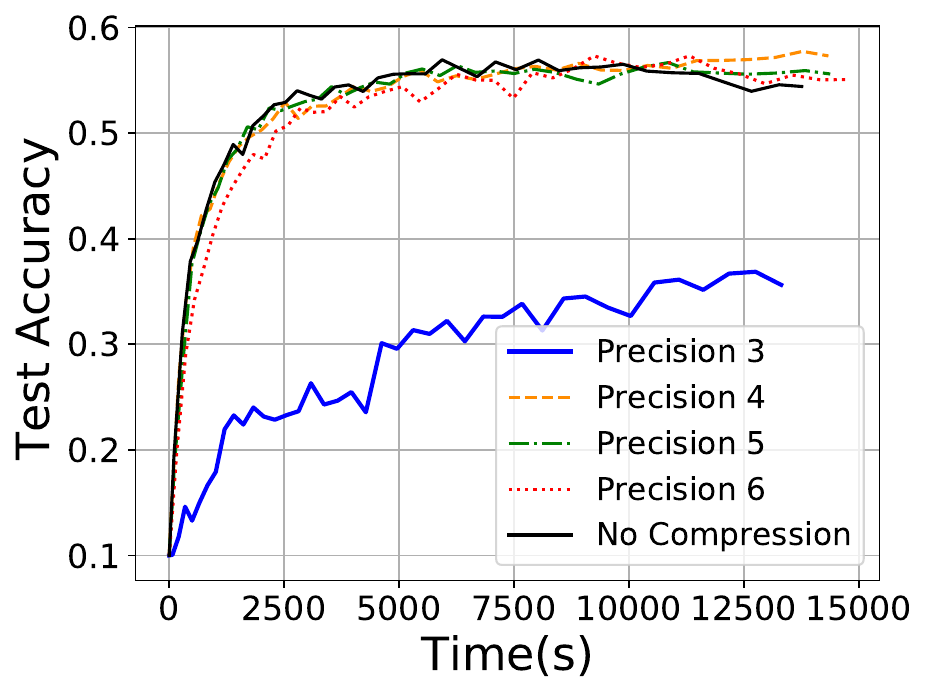}}
  \subfigure[Uploaded data vs. accuracy (the X-axis is in log-scale).]{\includegraphics[scale=0.27]{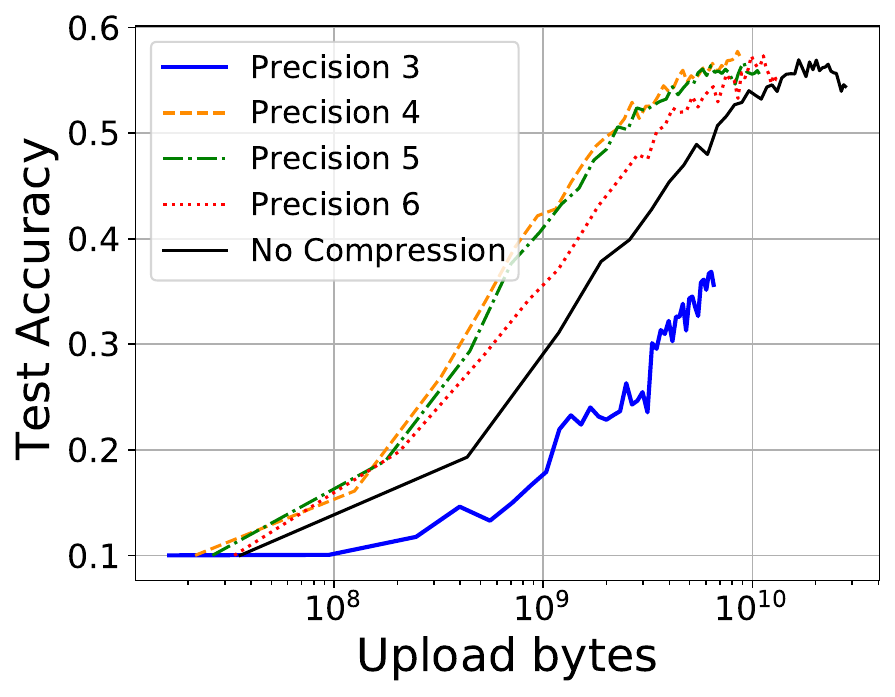}}
      \vspace{-1em}

  \caption{
  Impact of {\model}'s compression precision on the prediction performance and the communication cost, for the 
  CIFAR-10 Non-i.i.d. 2-class dataset. All results are plotted with the average of every $40$ global rounds.
  }\label{compress-degree}
      \vspace{-1em}

\end{figure}

\subsection{Communication Efficiency}
\label{subsec:compression_eval}

\subsubsection{Communication Cost.}
We next evaluate the network communication cost of {\model} in terms of the amount of data transferred via network. Table~\ref{compress-byte} shows the amounts of data transferred between the clients and the server (i.e., counting both model uploading and downloading) in order to achieve the target accuracy. FedAsync incurs the highest communication cost -- about $9.5\times$ of {\model}, and is not able to reach the target prediction performance. This confirms that severe communication bottleneck problem exists in asynchronous FL methods, where the server simply communicates with all the clients. FedAvg and TiFL have similar communication cost as they both use the same synchronous updating mechanism. {\model} incurs the lowest communication cost with compression technique and the proposed weighted aggregation on server.

Figure~\ref{compress-curve} further compares the uploaded bytes (from clients to the server) needed to reach a certain test accuracy. To achieve a relatively higher accuracy, 
{\model} needs fewer bytes than all the other three FL methods. More importantly, to achieve the same prediction performance for the CIFAR-10 2-class Non-i.i.d. dataset, {\model} requires up to 1.28$\times$ less data uploaded to the server, again demonstrating the efficiency and effectiveness of the model compression method used by {\model}.

\subsubsection{The Accuracy vs. Communication-Cost Tradeoff.}
\label{subsubsec:compression}

Next, we explore the accuracy vs. communication-cost tradeoff by varying the precision of {\model}'s compressor. {\small\texttt{Precision 3}} (i.e., a precision of three decimal places)
leads to the worst prediction performance, as shown in Figure~\ref{compress-degree}. This is because compressing the model by keeping only three digits after the decimal loses much information that is needed to converge the model; as a result, more training rounds, and more data communication, are needed in order to achieve a desirable accuracy. {\small\texttt{Precision 4}} is robust enough to strike a balance between the prediction performance and communication efficiency. {\small\texttt{Precision 4}} approaches the optimal accuracy achieved when no compression is used (Figure~\ref{compress-degree}(a)), while effectively reducing the amount of data uploaded by 36.41\% and 67.3\% (given the same target accuracy of 50\%) compared to {\small\texttt{Precision 6}} and {\small\texttt{No Compression}}, respectively (Figure~\ref{compress-degree}(b)). 
{\model} achieves a compression ratio (i.e., the ratio of the data size after compression and before compression) up to $3.5\times$ overall. 
{\model} is configured to use {\small\texttt{Precision 4}} as the default compression configuration in all the other experiments.

\begin{figure}[t]
  \centering
   \includegraphics[width=.4\textwidth]{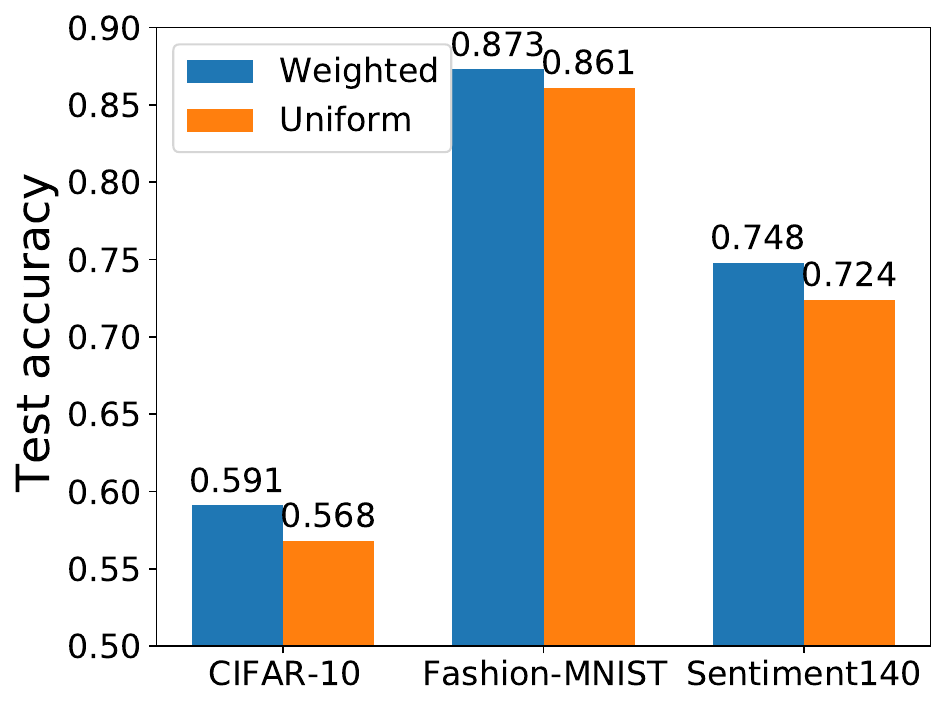}
  \caption{Comparison of {\model}'s weighted aggregation heuristic vs. a uniform baseline that assigns uniform weights when aggregating models from different tiers. 
  }\label{average}
\end{figure}

\subsection{Effectiveness of Weighted Aggregation}\label{com-aggregation}

We validate the effectiveness of our weighted aggregation heuristic. Weighted aggregation assigns more weight to the tiers that participate in the global training less frequently to prevent training bias towards the faster tiers. As shown in Figure~\ref{average}, the weighted aggregation heuristic improves the best test accuracy by 1.39\% to 4.05\%, compared to the baseline case, for the three datasets, demonstrating the effectiveness of the proposed approach.

\begin{figure}[t]
\centering
\subfigure[Accuracy over time.]{\includegraphics[scale=0.27,trim=0 0 0 5, clip]{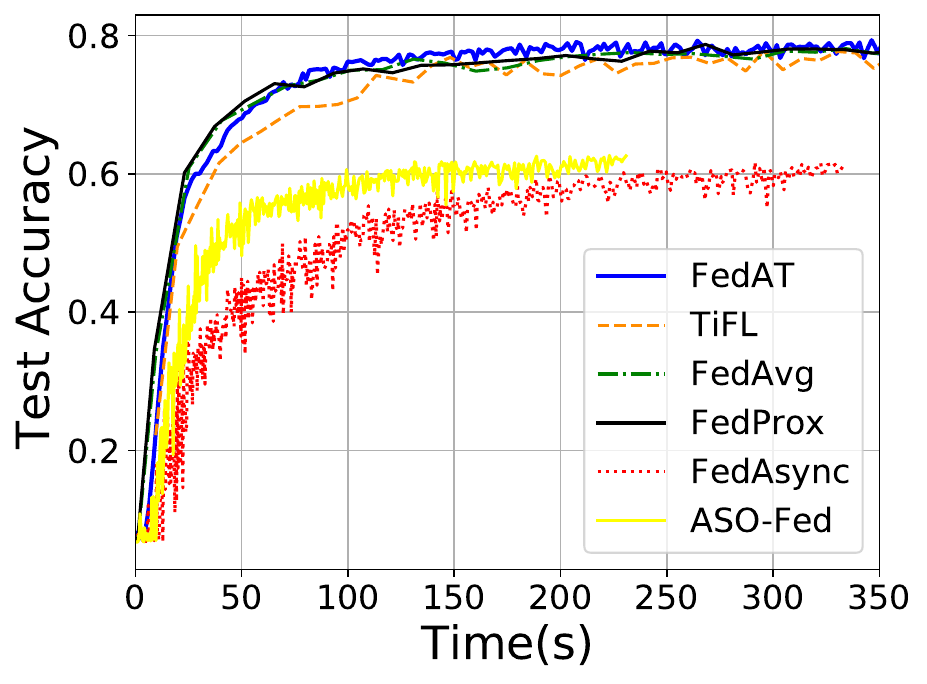}}
  \subfigure[Accuracy over uploaded bytes.]{\includegraphics[scale=0.27,trim=0 0 0 5, clip]{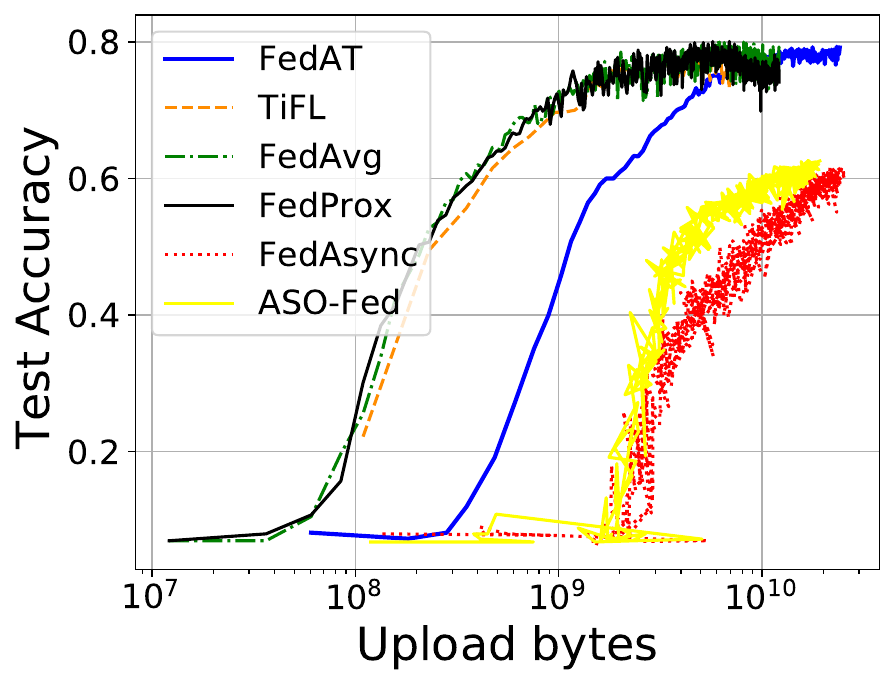}}    
\setcounter{subfigure}{0}
\vspace{-5pt}
\caption{Prediction accuracy of FEMNIST as a function of training time (a) and cumulative amount of data uploaded from clients to the server (b).}

\vspace{-5pt}
\label{fig:femnist}
\end{figure}

\vspace{-6pt}
\subsection{Large-Scale Training}
\label{subsec:large_scale}

In this test, we conduct large-scale experiments on the FEMNIST and Reddit datasets with $500$ participating clients deployed on 100 \emph{c5.2xlarge} AWS EC2 VMs. 

As shown in Figure~\ref{fig:femnist}(a), {\model} achieves the highest accuracy at the early stage of the training process, while maintaining at least $1.2\%$ higher accuracy than state-of-the-art synchronous methods, FedProx and TiFL. The two asynchronous FL methods, FedAsync and ASO-Fed, still perform worse than other synchronous methods. In addition, FedAsync and ASO-Fed see much higher communication cost than that of {\model}.

Although {\model} incurs higher communication cost than the synchronous FL methods at the early training stage due to asynchronous, cross-tier training, 
{\model} eventually achieves similar communication efficiency as the synchronous methods when these synchronous methods reach the highest prediction accuracy. This is because, with more frequent model update between tiers and the server, {\model} converges faster than synchronous baselines.

\begin{figure}[t]
\centering
\subfigure[Accuracy over time.]{\includegraphics[scale=0.27,trim=0 10 0 5, clip]{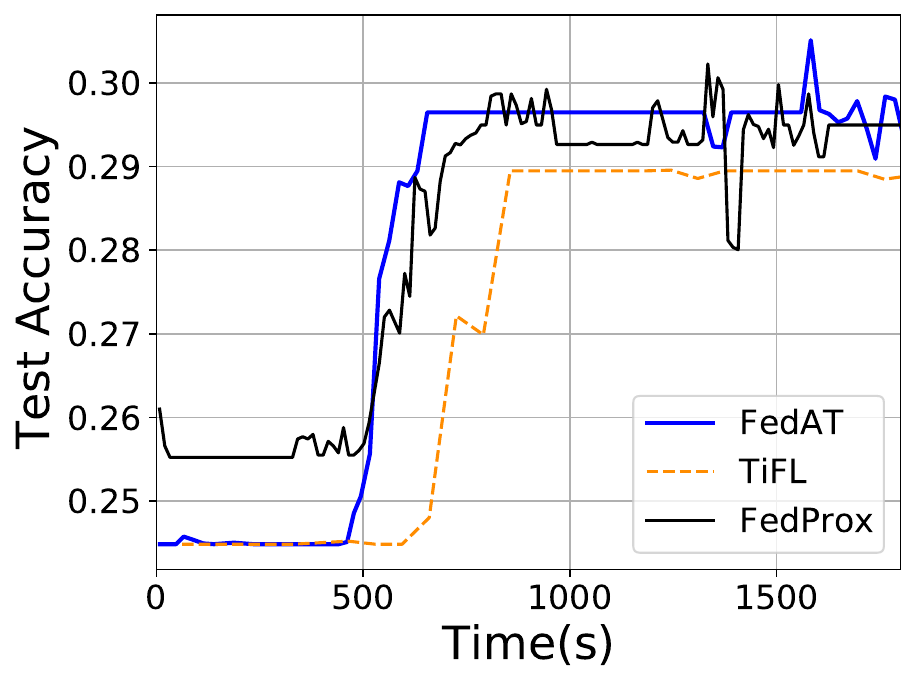}}
\subfigure[Loss over time.]{\includegraphics[scale=0.27,trim=0 10 0 5, clip]{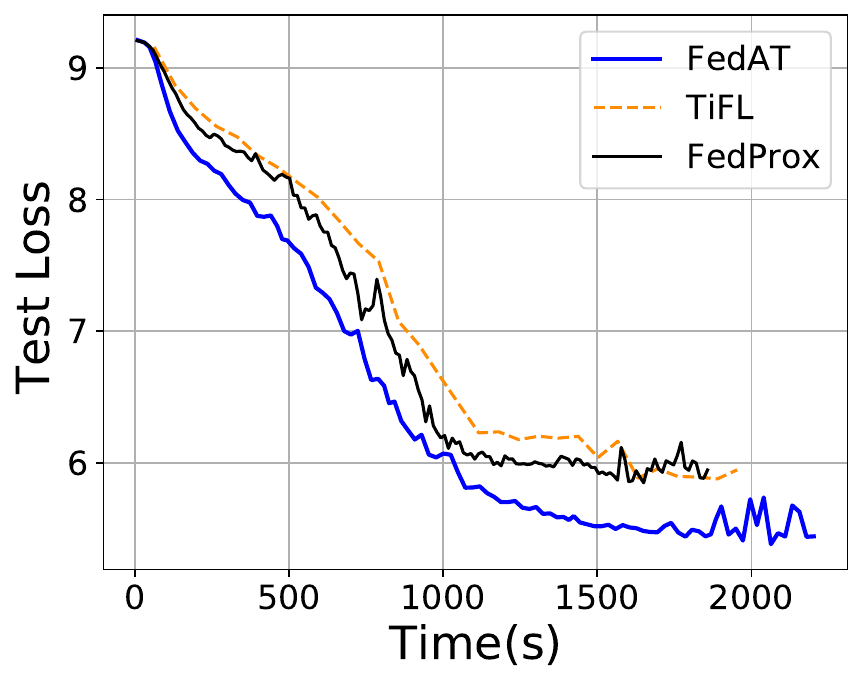}}    
\setcounter{subfigure}{0}
\vspace{-5pt}
\caption{Prediction accuracy (a) and loss (b) of Reddit as a function of training time.}

\vspace{-5pt}
\label{fig:femnist4}
\end{figure}

Figure~\ref{fig:femnist4} shows the prediction accuracy and loss on the Reddit dataset. Asynchronous FL baselines (FedAsync and ASO-Fed) have much lower prediction performance with no convergence trend on the Reddit dataset, therefore we omit their results in this test. 
We compare {\model} with TiFL and FedProx, which perform the best among all baseline FL methods. As shown in Figure~\ref{fig:femnist4}, we observe similar learning trend for the three frameworks, but {\model} has better prediction performance. Figure~\ref{fig:femnist4}(b) shows that {\model} achieves the lowest loss during the whole training process.

\begin{figure}[t]
\centering
\subfigure{
\includegraphics[scale=0.25,trim=10 80 10 80, clip]{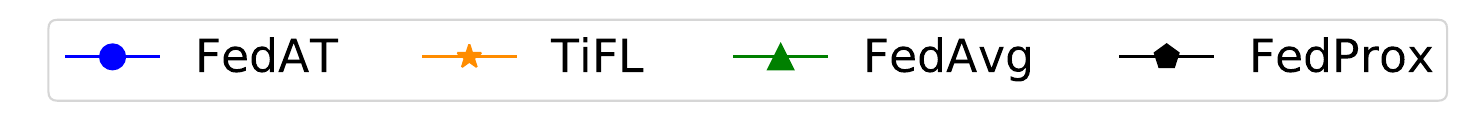}  
}\\
\setcounter{subfigure}{0}

\subfigure[CIFAR-10 Non-i.i.d.(2).]{\includegraphics[scale=0.26,trim=0 0 0 5, clip]{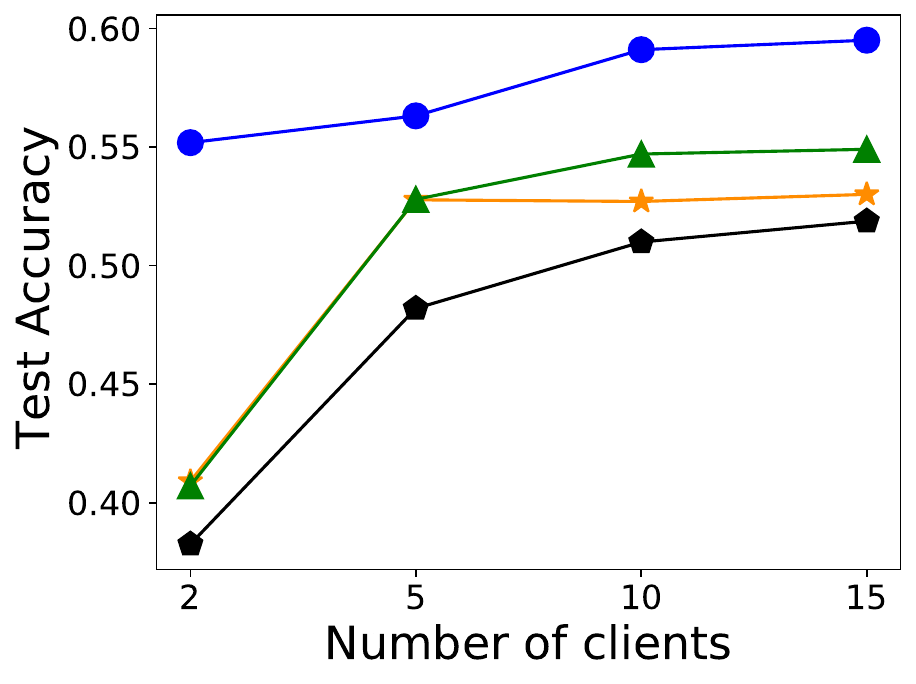}}
  \subfigure[Sentiment140.]{\includegraphics[scale=0.26,trim=0 0 0 5, clip]{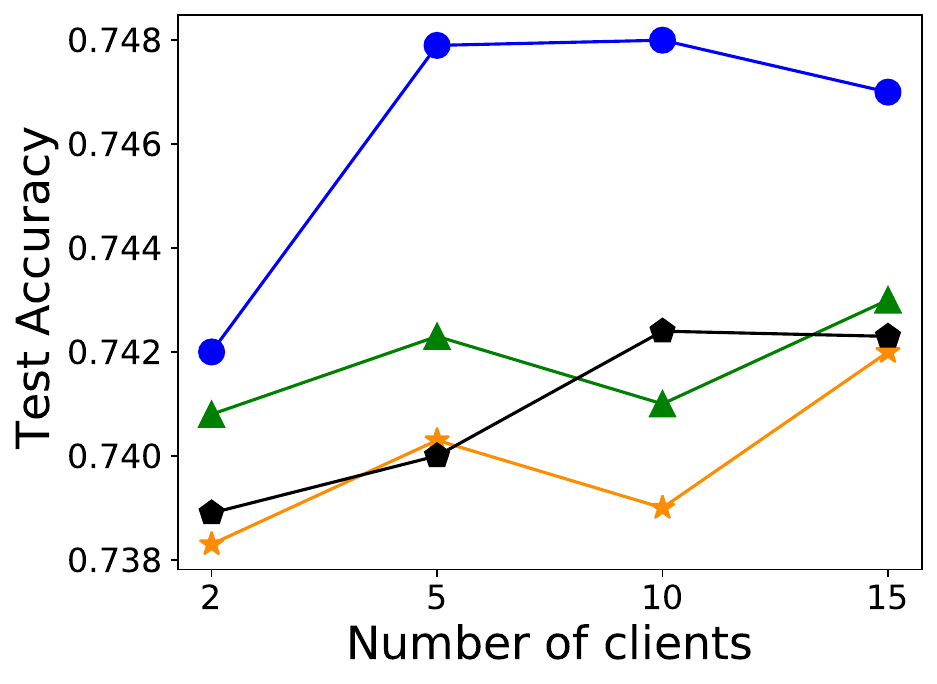}}    
\vspace{-10pt}
\caption{Prediction accuracy of CIFAR-10 (left) and Sentiment140 (right) as a function of the number of 
participating clients in one iteration. This test compares {\model} against three other FL methods (FedAvg, TiFL, FedProx), which all feature synchronous updating. }
    \vspace{-5pt}

\label{client-number}
\end{figure}

\subsection{Sensitivity Analysis}

\subsubsection{Impact of Client Participation Level.}\label{sec:part_level}

We next conduct a sensitivity study to quantify the impact of client participation level on the training accuracy. 
In a real-world situation,  for the communication efficiency consideration, it is often desirable to have as few clients as possible that participate in each global iteration.
In FedAvg, if a non-representative subset of clients is selected, the optimization process can deviate away from the minimum, which might lead to catastrophic forgetting~\cite{goodfellow2013empirical}. 
Although TiFL adopts the same tiering strategy as \model, TiFL achieves a similar performance as FedAvg. This is because,
as discussed in \cref{sec:stragglers}, TiFL uses the same synchronous update scheme as FedAvg, and tiering inevitably slows down the convergence speed but may not affect the final prediction accuracy when the training eventually converges. 

We observe in Figure~\ref{client-number} that reducing the level of client participation has negative effect on all of the four FL methods. \model~is robust in the non-i.i.d. case, where the prediction accuracy slightly decreases when the number of client participation decreases. While partial participation may reduce the convergence speed of \model~, the optimization can still achieve an optimal solution with the local constraint term. 
\model~suffers much less from a reduced participation level than FedAvg and TiFL. Even in the extreme case where only 2 out of 100 clients participate in each round of training, \model~still achieves $14.47\%$, $14.28\%$ and $16.93\%$ higher accuracy than FedAvg, TiFL, and FedProx on CIFAR-10, respectively. This is because the asynchronous, cross-tier training allows more clients to contribute to the global model, thus increasing the test accuracy on all clients.

\begin{figure}[t]
\centering
\includegraphics[width=.4\textwidth]{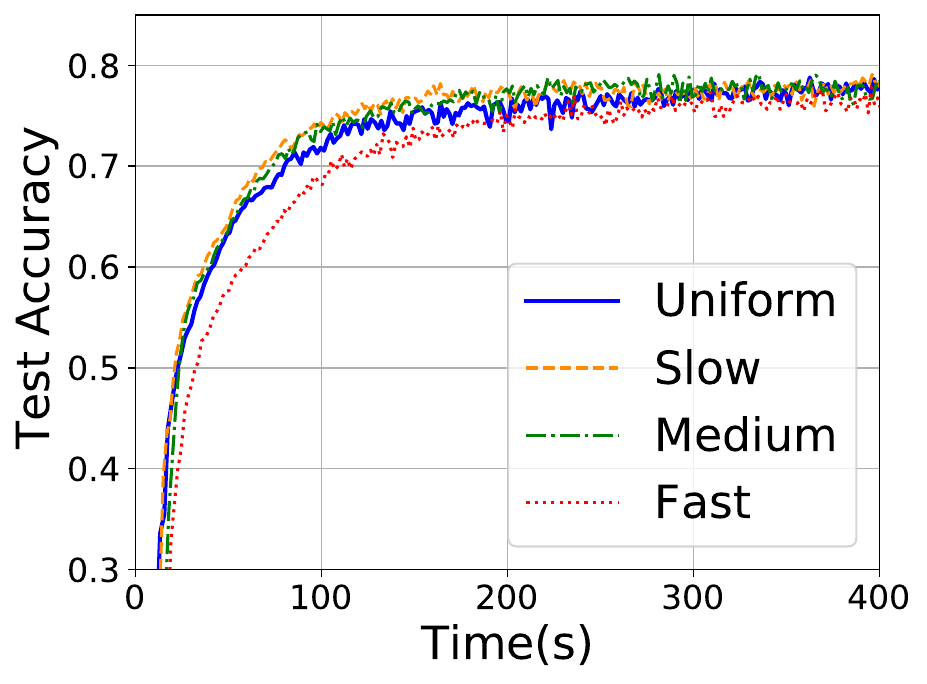}
\vspace{-5pt}
\caption{
Comparison of prediction accuracy over time on FEMNIST under different configurations of client distribution across tiers.}

\vspace{-5pt}
\label{fig:femnist2}
\end{figure}

\subsubsection{Impact of Number of Clients in Tiers.}

We next evaluate the robustness and resilience of {\model} to changing number of clients in different tiers. We partition a total of $500$ clients to five performance tiers, where each tier either has good resources or less amount of training data.
Specially, we test the following four configurations:
\emph{Uniform}: the baseline configuration that assigns the same number of clients to each tier with a distribution of $100/100/100/100/100$.
\emph{Slow}: where the slowest tiers (i.e., Tier 5) has the largest number of clients with a distribution of $50/50/100/100/200$.
\emph{Medium}: where the medium tier (i.e., Tier 3) gets the largest number of clients with a distribution of $50/100/200/100/50$.
\emph{Fast}: where the fastest tier (i.e., Tier 1) has the largest number of clients with a distribution of $200/100/100/50/50$. 
As shown in Figure~\ref{fig:femnist2}, all the four partition configurations eventually converge with close prediction performance, thanks to {\model}'s hybrid, synchronous, intra-tier training and asynchronous, cross-tier training strategy. 
\emph{Slow} and \emph{Medium} converges slightly faster than \emph{Fast}, since clients in \emph{Fast} configuration either have good resources or less amount of data.
As a result, training for the same number of rounds yield less accurate model in \emph{Fast} configuration as we are training on overall less amount of data. 
The results indicate that varying the tier sizes
would affect the convergence speed marginally but would not impact the model performance eventually when the training converges.


\section{Conclusion}

We have presented {\model}, a new FL method that maximizes the prediction performance and minimizes the communication cost using a tiered, hybrid synchronous-asynchronous training mechanism. {\model} cohesively synthesizes the following modules: 
(1) a tiering strategy to handle stragglers;
(2) an asynchronous scheme to update the global model among tiers for enhanced prediction performance;
(3) a novel, weighted aggregation heuristic that the FL server uses to balance the model parameters from heterogeneous, straggling tiers;
and (4) a polyline-encoding-based compression algorithm to minimize the communication cost. 
We have provided rigorous theoretical analysis for our proposed method for two general classes of convex and non-convex losses. We show that {\model} has provable model performance guarantee. Our evaluation has empirically validated our theoretical analysis, and demonstrates that {\model} achieves the highest prediction performance, converges the fastest, and is communication-efficient, compared to state-of-the-art FL methods.


\section*{Acknowledgments}

We are grateful to the anonymous reviewers for their valuable feedback and suggestions that improved the paper. This work is sponsored in part by the National Science Foundation (NSF) under CCF-1919075, CCF-1919113, CMMI-2134689, IIS-1755850, CNS-1841520, IIS-2007716, OAC-2007976, IIS-1942594, IIS-1907805, a Jeffress Memorial Trust Award, Amazon Research Award, NVIDIA GPU Grant, and Design Knowledge Company (subcontract number: 10827.002.120.04).
Part of the results presented in this paper were obtained using the Chameleon testbed supported by NSF.

\bibliographystyle{acm}
\bibliography{sample-base}

\newpage
\appendix
\onecolumn
\begin{center}
    \huge \textbf{Appendix}
\end{center}
\section{Theoretical Analysis of \model}
We analyze \model~in the setting of both convex and non-convex situations in this section. 

Recall that $w^t$ is the model parameters of the server maintained at the $t$-th round. Let $\bar g_t(w^t)  = \sum_{k=1}^{c} \frac{n_k}{N_c} \nabla h_k(w^t)$, in which, $N_c$ is the total number of data samples across all $c$ clients at tier $m$. Therefore, $w^{t+1} = w^t - \frac{T_{tier_{(M+1-m)}}}{T}\eta \bar g_t(w^t)$.

\subsection{Proof of Lemma \ref{bound_g}}\label{lemma1_proof}
 
To prove Theorem \ref{convex_theo}, we first introduce two Lemmas.

\begin{proof}
Using the notion of $\gamma$-inexactness for each local objective. We have
\begin{equation}
      \nabla h_k(w^{t}) = F_k(w^{t}) + \lambda (w^{t} - w_0) 
\end{equation}
 \begin{equation}\label{inexact}
      ||\nabla h_k(w^{t})|| \leq \gamma ||\nabla F_k(w^t)||
 \end{equation}

With $\bar g_t(w^t) = \sum_{k=1}^{c} \frac{n_k}{N_c} \nabla h_k(w^t)$,
we can get
 \begin{equation}
 \begin{aligned}
     ||\bar g_t(w^t)||^2 &=
     \frac{1}{N_c^2} ||n_1\nabla h_1(w^t) + n_2 \nabla h_2(w^t) +, ..., + n_c \nabla h_c(w^t)||^2 \\
     & \leq \frac{1}{N_c^2}\cdot N_c^2 ||\nabla h_1(w^t) +  \nabla h_2(w^t) +, ..., + \nabla h_c(w^t)||^2 \\
     & \leq m^2 ||\nabla h_{k^*}(w^t)||^2 \ \ \ (k^*=\arg\max_k \nabla h_k(w^t))\\ 
     & \leq m^2 \gamma^2 ||\nabla F_{k^*}(w^t)||^2 \;\;\; (with\; Eq.(\ref{inexact})) 
 \end{aligned}
 \end{equation}
 Take expectation of both sides and with Assumption \ref{bound_gradient}, we have
 \begin{equation}
 \begin{aligned}
     \mathbb{E}||\bar g_t (w^t)||^2 
     & \leq m^2 \gamma^2 \mathbb{E}||\nabla F_{k^*}(w^t)||^2 \\
     & \leq \gamma^2 G^2 c^2
 \end{aligned}
 \end{equation}
 
\end{proof}

\subsection{Proof of Lemma \ref{lemma_2}}\label{lemma2_proof}

\begin{proof}
$f(w)$ is $\mu$\textit{-strongly convex}, we can get:
\begin{equation}
f(w') - f(w^t) \geq \langle \nabla f(w^t), w' - w^t \rangle + \frac{\mu}{2}||w'-w^t||^2,
\end{equation}
Let us define $\Gamma(w')$ such that:
\begin{equation}
\Gamma(w') = f(w^t) + \langle \nabla f(w^t), w' - w^t \rangle + \frac{\mu}{2} ||w'-w^t||^2,
\end{equation}
$\Gamma(w')$ is a quadratic function of $w'$, then it has minimal value when $\nabla \Gamma(w') = \nabla f(w^t) + \mu(w'-w^t) = 0$. Then the minimal value of $\Gamma(w')$ is obtained when $w' = w^t - \frac{\nabla f(w^t)}{\mu}$, which is:
\begin{equation}
\Gamma_{\text{min}} = f(w^t) - \frac{||\nabla f(w^t)||^2}{2\mu}, 
\end{equation}
For $f(w)$ is $\mu$-\textit{strongly convex}, we can complete the proof: 
\begin{equation}
f(w_*) \geq \Gamma(w_*) \geq \Gamma_{\text{min}} = f(w^t) -\frac{||\nabla f(w^t)|| ^2}{2\mu},
\end{equation}
\begin{equation}
2\mu(f(w^t) - f(w_*)) \leq  ||\nabla f(w^t)||^2.
\end{equation}
\end{proof}

\subsection{Proof of Theorem \ref{convex_theo}}\label{convex_proof}
\begin{proof}
Now we start to prove the convergence of Theorem \ref{convex_theo}.
With Definition \ref{smoothness} we can get:
\begin{equation}\label{convex_eq1}
\begin{aligned}
    &f(w^{t+1}) - f(w^t)\\ 
    &\leq \langle \nabla f(w^t), w^{t+1} - w^t \rangle + \frac{L}{2} ||w^{t+1} - w^t||^2 \qquad(f(\cdot) \;is\; \textit{L-smooth})\\
    & = -\nabla f(w^t)^{\top} \frac{T_{tier_{(M+1-m)}}}{T}\eta \bar g_t(w^t) +  \frac{L\eta^2}{2}\frac{T_{tier_{(M+1-m)}}^2}{T^2}||\bar g_t(w^t)||^2  \qquad ( w^{t+1} = w^t - \frac{T_{tier_{(M+1-m)}}}{T}\eta \bar g_t(w^t))  
\end{aligned}
\end{equation}

Let $B = \frac{T_{tier_{(M+1-m)}}}{T}$. Then with Lemma \ref{bound_g}, we can update Equation \ref{convex_eq1} as
\begin{equation}\label{convex_eq2}
\begin{aligned}
    &\mathbb{E}[f(w^{t+1})] - f(w^t) \\
    &\leq -\nabla f(w^t)^{\top}B\eta \mathbb{E}[\bar g_t(w^t)] + \frac{L}{2} \eta^2 B^2 \mathbb{E}||\bar g_t(w^t)||^2\\
    &\leq -\nabla f(w^t)^{\top}B\eta \mathbb{E}[\bar g_t(w^t)] + \frac{L}{2} \eta^2 \gamma^2 B^2 G^2 c^2
\end{aligned}
\end{equation}

Then from Assumption \ref{bound_gradient2}, we have
\begin{equation}\label{convex_eq3}
\begin{aligned}
    &\mathbb{E}[f(w^{t+1})] - f(w^t) \\
    &\leq -B\eta \sigma||\nabla f(w^t)||^2 + \frac{L}{2} \eta^2 \gamma^2 B^2 G^2 c^2 
\end{aligned}
\end{equation}

Then with Lemma \ref{lemma_2}, Equation (\ref{convex_eq3}) can be updated as
\begin{equation}\label{convex_eq3.1}
\begin{aligned}
    &\mathbb{E}[f(w^{t+1})] - f(w^t) \\
    & \leq -2\mu B \eta \sigma (f(w^t) - f(w_*)) + \frac{L}{2} \eta^2 \gamma^2 B^2 G^2 c^2
\end{aligned}
\end{equation}

By subtracting $f(w_*)$ from both sides and moving $f(w^t)$ from left to right, we get
\begin{equation}\label{convex_eq4}
\begin{aligned}
    &\mathbb{E}[f(w^{t+1})] - f(w_*) \\
    & \leq -2\mu B \eta \sigma (f(w^t) - f(w_*)) + (f(w_t) - f(w_*)) + \frac{L}{2} \eta^2 \gamma^2 B^2 G^2 c^2 \\
    &= (1-2\mu B \eta \sigma)(f(w^t)-f(w_*))+ \frac{L}{2} \eta^2 \gamma^2 B^2 G^2 c^2 
\end{aligned}
\end{equation}

Taking the whole expectations and rearranging (\ref{convex_eq4}), we obtain
\begin{equation}\label{convex_eq5}
\begin{aligned}
    &\mathbb{E}[f(w^{t+1}) - f(w_*)] \\
    &\leq (1-2\mu B \eta \sigma)\mathbb{E}[(f(w^t)-f(w_*))]+ \frac{L}{2} \eta^2 \gamma^2 B^2 G^2 c^2 
\end{aligned}
\end{equation}
subtracting $\frac{L\eta \gamma^2 B G^2 m^2}{4\mu \sigma}$ from both sides, we have
\begin{equation}\label{convex_eq6}
\begin{aligned}
    &\mathbb{E}[f(w^{t+1}) - f(w_*)] - \frac{L\eta \gamma^2 B G^2 c^2}{4\mu \sigma}\\
    &\leq (1-2\mu B \eta \sigma)(\mathbb{E}[(f(w^t)-f(w_*))] - \frac{L\eta \gamma^2 B G^2 c^2}{4\mu \sigma}  )
\end{aligned}
\end{equation}
The left side of (\ref{convex_eq6}) is a geometric series with common ratio $1-2\mu B \eta \sigma$, when $t+1 = T$, we get Equation (\ref{convex_con}), then we complete the proof.
\end{proof}

\subsection{Proof of Theorem \ref{nonconvex_theo}}
\begin{proof}
Take expectation at both sides of Equation (\ref{convex_eq3}), we have
\begin{equation}\label{nonconvex_eq1}
\begin{aligned}
    &\mathbb{E}[f(w^{t+1})] - \mathbb{E}[f(w^t)] \\
    &\leq -B\eta \sigma\mathbb{E}[||\nabla f(w_t)||^2] + \frac{L}{2} \eta^2 \gamma^2 B^2 G^2 c^2 
\end{aligned}
\end{equation}

Then sum Equation (\ref{nonconvex_eq1}) at both sides over global iteration $T$. We have
\begin{equation}\label{nonconvex_eq2}
\begin{aligned}
    &\mathbb{E}[f(w^{t+1})] - f(w^0) \\
    &\leq \sum_{t=0}^{T-1}-B\eta \sigma \mathbb{E}[||\nabla f(w_t)||^2] + \frac{L}{2} T^2\eta^2 \gamma^2 B^2 G^2 c^2
\end{aligned}
\end{equation}

As $\min f(w^t) = f(w_*)\leq \mathbb{E}[f(w^{t+1})]$, then we have
\begin{equation}\label{nonconvex_eq2_5}
\begin{aligned}
    f(w_*) 
    &\leq f(w^0) -\sum_{t=0}^{T-1}B\eta \sigma \mathbb{E}[||\nabla f(w_t)||^2]\\ &+ \frac{L}{2} T^2\eta^2 \gamma^2 B^2 G^2 c^2
\end{aligned}
\end{equation}
Rearrange (\ref{nonconvex_eq2_5}) we can get
\begin{equation}\label{nonconvex_eq3}
\begin{aligned}
    &\sum_{t=0}^{T-1} B \mathbb{E}[||\nabla f(w_t)||^2] \\ 
    &\leq \frac{f(w^0) -f(w_*)}{B\eta \sigma} + \frac{L}{2\sigma} T^2\eta \gamma^2 B G^2 c^2
\end{aligned}
\end{equation}
Then we complete the proof.
\end{proof}

\end{document}